\documentclass[hyper, a4paper, 11pt, oneside]{JHEP3}
\usepackage{epsfig}
\usepackage{amsmath,amsfonts,amssymb}
\title{Structure formation in $f(R)$ gravity: A distinguishing probe between the dark energy
and modified gravity}

\author{ Shant Baghram and Sohrab Rahvar \\ Department of Physics, Sharif University of
Technology, P.O.Box 11365--9161, Tehran, Iran\\
Email:baghram@physics.sharif.edu , rahvar@sharif.edu}

\abstract {In this work, we study the large scale structure
formation in the modified gravity in the framework of Palatini
formalism and compare the results with the equivalent smooth dark
energy models as a tool to distinguish between these models. Through
the inverse method, we reconstruct the dynamics of universe,
modified gravity action and the structure formation indicators like
the screened mass function and gravitational slip parameter.
Consequently, we extract the matter density power spectrum for these
two models in the linear regime and show that the modified gravity
and dark energy models predictions are slightly different from each
other at large scales. It is also shown that the growth index in the
modified gravity unlike to the dark energy models is a scale
dependent parameter. We also compare the results with those from the
modified gravity in the metric formalism. The modification on the structure formation can
also change the CMB spectrum at large scales however due to the
cosmic variance it is hard to detect this signature. We show that a
large number of SNIa data in the order of 2000 will enable us to
reconstruct the modified gravity action with a suitable confidence
level and test the cosmic acceleration models by the structure
formation.}

\begin{document}
\section{Introduction}

For more than a decade, the positive acceleration of the universe is
one of the challenging questions in  physics \cite{R04,Dun09}. The
physics and the mechanism behind this phenomenon is unknown. The
Cosmological Constant (CC) is the most straightforward suggestion to
describe an accelerating universe \cite{Adam98,Davis07}. However the
fundamental questions like the fine tuning and coincidence problems
opened new horizons to introduce the alternative models
\cite{Wienberg89} like Dark Energy (DE) \cite{Peeb03} and models of
the Modified Gravity (MG) \cite{Carroll04}. The other motivation for
the modified gravity models is unifying the dark matter and the dark
energy problems in a unified formalism \cite{rahsaff}.

The MG theories need to be tested in two domains of the
observations: (a) cosmological scales \cite{Jain08} and (b) local
gravity \cite{Faraoni07}. The simplest modification of gravity is
the extension of Einstein-Hilbert action where instead of Ricci
scalar, we use a generic function of $f(R)$. These modified gravity
theories are divided into the Metric and Palatini formalism
\cite{Tsujikawa2010}. There are a lot of debates in the literature
on the viability of this type of modified gravity model concerning
the local gravity tests \cite{Faraoni04} where a vast range of them
had been falsified \cite{Barausse2008,Kobayashi2008}. Also many
models are proposed to evade the falsification criteria
\cite{Upadhye2009,HuSawicki2007,Starobinsky}. The Palatini formalism
is a second order differential equation which has more simpler
structure than the fourth order metric formalism. Since all the
conventional  field equations  are described by the second order
differential equation, it seems that gravity should also follow this
rule. So this is the main advantage of working with the Palatini
formalism. From cosmological point of view, we will show that for a
given dynamics of the Universe there is a one-to-one map to the
corresponding action in the Palatini formalism. Also in this
formalism we don't have the instability problem as in the metric
formalism. While there are advantages in this formalism, there is
challenging questions in the Palatini formalism as the microscopic
behavior of the matter fields where in the Einstein frame, these
models disagrees with the standard scalar-matter coupling theories
\cite{Flanagan2004}. Means that we can not apply the perturbation
theory in these scales as the gravity field is directly relates to
the energy-momentum tensor, in contrast to the Einstein gravity
where the amount of the perturbation of the metric at a given point
is averaged over all the space. Although there are debates on how we
should do averaging procedure over the microscopic scales \cite{Li},
it has been shown that the problematic microscopic behavior of this
theory can be solved in the case of $f(R)$ very near to the
$\Lambda$CDM model \cite{Tsujikawa2010}.


Keeping in mind that there are many fundamental questions in the
Palatini formalism , here in this work we mainly focus on the
behaviors of the modified gravity models at the cosmological scales
and their predictions on the structure formation. The main question
we address in this work is the investigation of cosmological probes
in the structure formation to distinguish between CC, DE and MG
models \cite{Bertschi2008}. We first reconstruct the dynamics of the
background and the corresponding MG action via the inverse method
from the SNIa data, described in \cite{Baghram09}. The dynamics of
the background (i.e Hubble parameter $H=H(z)$) can only distinguish
between CC and the alternative models \cite{Baghram09} however in
the level of probing the expansion history of the Universe, MG and
DE models are not distinguishable. Consequently, we use the
structure formation probes as the promising tool to distinguish
between alternative models \cite{Pogosian2008,Lue2004}. It is worth
to mention that the alternative models can be extended to a
complicated forms such as interacting dark energy-dark matter models
\cite{Wei2008} or clustering DE theories,  which can not be
distinguished from the MG models even in the level of the structure
formation probes \cite{Kunz2007}.

In this work we focus on the structure formation issue in the
modified gravity--Palatini formalism and compare the observational
effects with that in the smooth dark energy models (sDE). We derive
the screened mass function defined as the fraction of effective
gravitational constant to the standard gravitational theory as
$Q\equiv {G_{eff}}/{G_{N}}$. The effective gravitational constant
appears in the Poisson equation depends on the scale of the
structure. The other relevant parameter appear in our calculation is
the gravitational slip parameter $\gamma=-{\Phi}/{\Psi}$, which is
the fraction of spatial perturbation of the metric to the
perturbation of the time-time element. The screened mass function as
well as the gravitational slip parameter in DE models are equal to
one in GR while in MG models they are time and scale dependent
parameters  . This scale dependance of structure formation in
modified gravity theories is a well known effect discussed in
literature \cite{HuSawicki2007,Starobinsky,Zhang08}. A combination
of screened mass and gravitational slip parameters recently have
been used for distinguishing the MG in metric formalism with the
alternative models via analyzing the structure formation and weak
lensing \cite{rey2010}. We show that the derived reconstructed
matter density power spectrum and growth index in this model are
affected by the screened mass function. Also we study the effect of
gravitational slip parameter on the Integrated Sachs-Wolf (ISW) and
compare it with recent observations of WMAP data. As a comparison
with the metric formalism, we compare our results with that of
structure formation in the metric formalism.

The structure of this article is as follows: In section
\ref{section2} we review the standard equations governing the
relativistic structure formation. In section \ref{section3} we
re-derive the structure formation equations for DE and Palatini MG
models. Also in this section the metric formalism structure
formation equations are derived for comparison with that in the
Palatini formalism.
In section \ref{section4} we extract the screened mass function and
gravitational slip parameters, reconstructed from a modified gravity
equivalent to a dark energy model and the effect of screened mass
function on the power spectrum of the structures as well as the
growth index is studied. Also the screened mass in the metric
formalism is obtained in order to show the scale dependence of
structure's growth. Details of the calculation in metric formalism
is given in Appendix A. In section \ref{section5} we investigate the
effect of gravitational slip parameter on CMB power spectrum due to
the ISW effect and compare the results with recent WMAP data. The
conclusion is presented in section \ref{conc}.

\section{Structure Formation: Perturbation theory}

\label{section2} The relativistic structure formation in the
universe can be studied by the perturbation of the FRW metric and
homogenous energy-momentum tensor of the cosmic fluid \cite{mukh92}.
A standard technic is decomposition of the  metric and energy
momentum tensor perturbations into the scalar, vector and tensor
modes where each mode evolves independently \cite{Dodelson}. Since
in this work we only deal with the density contrast evolution of the
structures, we study the evolution of the scalar modes in our study.
One of the ambiguities in the perturbation of metric is the gauge
freedom, similar to what appears in the theory of Electromagnetism.
In fact the problem arises due to the excess number of the degrees
of freedom (i.e. number of variables) compare to the number of
independent field equations. We fix this freedom with so-called
Newtonian gauge as follows \cite{Ma1995}:
\begin{equation}
\label{EqNewtonian}
ds^2=-[1+2\Psi(\vec{x},t)]dt^{2}+a^{2}(t)[1+2\Phi(\vec{x},t)]g^{(3)}_{ij}.
\end{equation}
It should be noted that the two scalar perturbations of $\Psi$ and
$\Phi$ are gauge invariant variables and hence observable
parameters, as the electric and magnetic fields in the
electromagnetism \cite{Bardeen80}. In order to interpret the
physical meaning of the perturbations, let us look to the the
Newtonian limit. In the Newtonian limit, the geodesic equation
reduces to $\ddot{x}=-\vec{\nabla}\Psi(\vec{x},t)$. On the other
hand the Einstein equation in the weak field regime reduces to the
Poisson equation as follows $\nabla^2\Phi=4\pi G\rho\delta$.

The dynamics of the perturbations represent the evolution of the
structures which can be obtained from the Boltzman and the Einstein
field equations. First we consider the equations arise from the
perturbed field equations. The (0-0) component of perturbed Einstein
equation (i.e.$\delta G^{\mu}_{\nu}=\kappa \delta T^{\mu}_{\nu}$) in
the fourier mode is given by \cite{Dodelson}:
\begin{equation} \label{EqPoisson}
k^{2}\Phi+3H(\dot{\Phi}-H\Psi)=4\pi Ga^{2}\rho\delta,
\end{equation}
where derivative is in terms of the coordinate time defined in the
metric and we ignore the subscript $k$ for simplicity. On the other
hand the spatial component of the Einstein equation, after applying
the projection operator of $\hat{k}_i\hat{k}^j - 1/3 \delta_i^j$ in
Fourier space, results in:
\begin{equation} \label{EqShearGr}
k^{2}(\Phi+\Psi)=-32\pi G\rho^{0}_{r}\Theta_{2},
\end{equation}
where the right hand side of this equation is derived from the first
order perturbation of the distribution function of the cosmic
radiation. Here $\Theta_{2}$ is the the second moment of anisotropic
stress of radiation where $l$-st moment is defined by
\begin{equation}\label{EqThetal}
\Theta_{l}\equiv\frac{1}{(-i)^{l}}\int_{-1}^{+1}\frac{d
\mu}{2}P_{l}(\mu) \Theta(\mu),
\end{equation}
and  $P_{l}(\mu)$ is the Legendre polynomials. $\Theta(\mu)$ is the
temperature contrast of radiation and defined by:
\begin{equation} \label{EqTheta}
T(\vec{x},\hat{p}^i,t)=T_0[1+\Theta(\vec{x},\hat{p}^i,t)],
\end{equation}
where $T_0$ is the mean temperature of radiation, and the $\mu$ is the
cosine of the angle between the momentum of a fluid and the direction that
observer looks at the temperature contrast.

In the matter dominated universe where the energy density of
radiation is negligible, the right hand side of equation
(\ref{EqShearGr}) is zero, hence $\Psi=-\Phi$. In the next section
we will show that for the modified gravity models even in the matter
dominant epoch the scalar perturbations have no more simple relation
as in the general relativity. In order to discriminate the modified
gravity models from  sDE, we use the new parameter of $\gamma$,
so-called gravitational slip parameter as \cite{ame}:
\begin{equation}
\gamma=-\frac{\Phi}{\Psi},
\label{gamma}
\end{equation}
In the next section we relate the slip parameter to the observable
parameters in the structure formation. Any deviation from $\gamma=1$
can show the crackdown in our assumptions.

As we noted before, the other set of the equations governing the
structure formation are the perturbed Boltzman equations. For a
given perfect fluid for the universe with the energy momentum tensor
of $T_{\mu\nu}=(p+\rho)u_{\mu}u_{\nu}+pg_{\mu\nu}$, we can obtain
the energy and momentum conservation equations by
$T^{\mu\nu}_{~~~;\nu}=0$ for $\mu=0$ and $\mu=i$ as below:
\begin{equation}\label{Eqdc}
\frac{\dot{\delta}}{1+\omega}=-\frac{\theta}{a}-3\dot{\Phi}-3H(\frac{\delta p}{\delta \rho}-\omega)\frac{\delta}{1+\omega},
\end{equation}
\begin{equation}\label{EqEul}
\dot{\theta}=-H(1-3\omega)\theta-\frac{\dot{\omega}}{1+\omega}\theta+\frac{1}{a}\frac{\delta p}{\delta \rho}\frac{k^2\delta}{(1+\omega)}+\frac{k^2\Psi}{a},
\end{equation}
where $\theta$ is the divergence of peculiar velocity
$\theta=\nabla\cdot v$ and $\omega$ is the equation of state. Since
we assume that the cosmic fluids are already decoupled from each
other at the late time universe, equations (\ref{Eqdc}) and
(\ref{EqEul}) can be written for each component with the
corresponding equation of states.

The goal of structure formation theory is to find the dynamics of
the four perturbed quantities of $\Phi$ and $\Psi$ from the metric
and $\delta$ and $\theta$ from the energy momentum tensor through
the field and Boltzman Equations \cite{Padman}.

\section{Structure formation in accelerating universe}

\label{section3} In this section we reexpress the perturbation
equations for the structure formation in an accelerating universe.
We note that the derivation for sub-horizon scalar perturbations in
the general case of scalar-tensor gravities were obtained in
\cite{Boisseau} where the Palatini formalism is a special case of
it. In the first subsection we will introduce the equation of the
structure formation for a universe filled with a smooth dark energy
with an equation of state $\omega_{DE}=\frac{p_{DE}}{\rho_{DE}}<
-\frac{1}{3}$. In the second part we derive the equation of the
structure formation in the Palatini MG model equivalent to a Dark
Energy model.

\subsection{Smooth Dark Energy Models}
In this subsection, we study the structure formation assuming a DE
fills the universe. In a generic case, if we assume a clustering
dark energy model, the perturbation of Einstein equation in the
non-radiation dominate epoch can be written as \cite{Hu98}:
\begin{eqnarray}\label{EqPoissonQuint1}
\frac{k^2\Phi}{a^2}+3H(\dot{\Phi}-H\Psi)&=&4\pi
G[\rho_{m}\delta_{m}+\delta_{DE}\rho_{DE}]\\
k^2(\Phi+\Psi)&=&-32\pi G\sigma,
\end{eqnarray}
where $\sigma$ is the anisotropic stress of DE model. This quantity
is a generic property of a fluid with finite shear viscous
coefficients \cite{Schimd2006}. In case of our study, in which we
consider a perfect fluid for the matter and DE, the anisotropic
stress is vanished (i.e $\sigma=0$) and $\Phi=-\Psi$. Using the
trace of perturbed Einstein equation $\delta R = -8\pi G (3\delta p
- \delta\rho)$ and combining with Eq. (\ref{EqPoissonQuint1}), we
can eliminate the density perturbation and write a second order
differential equation for $\Phi$ as follows \cite{Malq2002}:
\begin{equation} \label{EqMoQuint1}
\ddot{\Phi}+4H\dot{\Phi}+(2\dot{H}+3H^2)\Phi=-4\pi G\delta p_{DE}.
\end{equation}
In the simple case where we don't have clustering of the DE we set
$\delta\rho_{DE}\approx 0$ and the dynamics of $\Phi$ can be
obtained from (\ref{EqMoQuint1}). We can substitute the numerical
value of $\Phi$ in (\ref{EqPoissonQuint1}) and obtain the evolution
of the density contrast for the matter as:
\begin{equation}
\frac{k^2}{a^2}\Phi+3H^2\Phi+3H\dot{\Phi}=4\pi G\delta \rho_{m}.
\label{diffphi}
\end{equation}
It is worth to mention that DE models with the clustering properties
can modify the structure formation and consequently change the
matter power spectrum \cite{Kunz2007}.

In order to compare our results in the following sections with the
observation, we reexpress the differential equation (\ref{diffphi})
in terms of the redshift:
\begin{equation} \label{EqQuiPhi}
\Phi^{ ' } - \frac {\Phi}{1+z} -\frac{1}{3}\frac{
\tilde{k}^2}{E(z)^2}(1+z)\Phi = -\frac{1}{2} (1+z)^2 \Omega^{0}_{m}
\delta_{m}.
\end{equation}
where $E(z)={H}/{H_{0}}$ is a dimensionless Hubble parameter,
$H_{0}$ is the Hubble parameter at the present time and $\tilde{k} =
{k}/{H_{0}}$ is the dimensionless wavenumber of the structures and
$\Omega_{m}^{0}$ is the matter density parameter at the present
time.

\subsection{ $f(R)$ Modified gravity theories in Palatini Formalism}

\label{subsection3-1} We take $f(R)$--modified gravity in the
Palatini formalism as an alternative model for the acceleration of
the universe. In this formalism the connection and metric are
independent fields. Variation of action with respect to these
fields, results in a set of second order differential equation for
the metric \cite{sotiriou09}.

The action of MG in palatini formalism is given by:
\begin{equation}
S[f;g,\hat{\Gamma},\Psi_{m}]=-\frac{1}{2\kappa}\int{d^{4}x\sqrt{-g}f(R)+
S_{m}[g_{\mu\nu},\Psi_{m}]},
\end{equation}
where $\kappa=8\pi{G}$ and $S_{m}[g_{\mu\nu},\Psi_{m}]$ is the
matter action depends on the metric $g_{\mu\nu}$ and the matter
fields $\Psi_{m}$, $R=g^{\mu\nu}{R_{\mu\nu}}(\hat{\Gamma})$ is the
generalized Ricci scalar and $R_{\mu\nu}$ is the Ricci tensor made
of affine connection. Varying the action with respect to the metric
results in
\begin{equation}
f'(R)R_{\mu\nu}(\hat{\Gamma})-\frac{1}{2}f(R)g_{\mu\nu}=\kappa
T_{\mu\nu}, \label{field}
\end{equation}
where prime is the derivative with respect to the Ricci scalar and
$T_{\mu\nu}$ is the energy momentum tensor defined by
\begin{equation}
T_{\mu\nu}=\frac{-2}{\sqrt{-g}}\frac{\delta(\sqrt{-g}{\cal
L}_m)}{\delta g^{\mu\nu}}.
\end{equation}
The trace of the field equation results in:
\begin{equation}
RF(R)-2f(R)=\kappa T, \label{EqPalTrace},
\end{equation}
where $T=g^{\mu\nu}T_{\mu\nu}=-\rho+3p$, and $F(R)={df}/{dR}$. First
of all we want to know the background evolution of the Universe in
Palatini formalism. We use FRW metric in flat universe (namely
$k=0$).
\begin{equation}
ds^{2}=-dt^{2}+a(t)^{2}\delta_{ij}dx^{i}dx^{j},
\end{equation}
and assume that universe is filled with a perfect fluid with the
energy-momentum tensor of $T^{\nu}{}_{\mu}=diag(-\rho,p,p,p)$, the
generalized FRW equations is as below \cite{Allemandi}:
\begin{equation}
H^{2}=\frac{1}{6(1-3\omega)f'}\frac{3(1+\omega)f-(1+3\omega)Rf'}
{\left[1+\frac{3}{2}(1+\omega)\frac{f''(2f-Rf')}{f'(Rf''-f')}\right]^{2}}.
\label{hpala}
\end{equation}
In the matter dominated epoch where $\omega = 0$, the Hubble
parameter reduces to:
\begin{equation} \label{EqPalHubble}
H^{2}=\frac{1}{6f'}\frac{3f-Rf'}{\left[1+\frac{3}{2}\frac{f''(2f-Rf')}{f'(Rf''-f')}\right]^{2}}.
\end{equation}

In order to study the  structure formation in MG models, we perturb
the metric in the same way as done in the previous part. In the
modified Einstein equation, we move the extra geometrical terms out
of the Einstein tensor to the right hand side of equation, recalling
it by ${T^{\mu}_{~\nu}}^{(dark)}$. The  perturbed  modified Einstein
equation can be written as follows:
\begin{equation}
\delta G^{\mu}_{~\nu}=\delta T^{\mu}_{~\nu}+\delta
T^{\mu(dark)}_{~\nu}. \label{pfe}
\end{equation}
The $(0,0)$ component of the field equation is
\cite{Tsujikawa08,Koiv2006}:
\begin{eqnarray}
\nonumber
&&3(2H+\frac{\dot{F}}{F})(H\Psi-\dot{\Phi})-\frac{2k^2}{a^2}\Phi+\frac{1}{F}(\frac{3\dot{F}^{2}}{2F}+3H\dot{F})\Psi\\
\nonumber &=&\frac{1}{F}\{-\kappa\delta\rho+\delta
F[3H^2-\frac{3\dot{F}^2}{4F^2}-\frac{R}{2}+\frac{k^2}{a^2}]\\
&+&(\frac{3}{2}\frac{\dot{F}}{F}+3H)\delta\dot{F} \},
\label{EqMoPoisson}
\end{eqnarray}
where $\delta F$ is the variation of the $F(R)$ due to the
perturbation of the metric. Eq.(\ref{EqMoPoisson}) is a modified
Poisson equation where for $F=1$ we recover equation
(\ref{EqPoissonQuint1}). $G_{i}^{0}$-component of field equation
also is given as below:
\begin{equation}\label{EqMoMoment}
H\Psi-\dot{\Phi}=\frac{1}{2F}\left[\delta\dot{F}-
(H+\frac{3\dot{F}}{2F})\delta F-\dot{F}\Psi+\kappa \rho_{m}
\frac{\theta_k}{k^2} \right],
\end{equation}
where $\theta_k = {k}^iT^0{}_i/(\rho+p)$ is a gauge invariant
quantity. Now we combine equation (\ref{EqMoPoisson}) and
(\ref{EqMoMoment}) to obtained the Poisson equation in the modified
gravity as \cite{Song10}:
\begin{equation} \label{EqScreenedMass}
k^{2}\Phi=4\pi Ga^{2}\Delta_{m}\rho Q(k,z),
\end{equation}
where
\begin{equation}
\Delta_{m}\equiv \delta_{N}+3H\theta_k/{k^2} \label{dc}
\end{equation}
is a gauge invariant quantity in terms of the density contrast
{\cite{Amendola2010}} and $Q(k,z)$ is so-called screened mass. This
term in the Palatini formalism is given as
\begin{equation}\label{EqPalScreenedMass}
Q=\frac{1}{F}(1+\frac{mR^{-1}D}{1-m}),
\end{equation}
where $m=R \frac{d}{dR} \ln F(R)$ is a dimensionless deviation
parameter and $D$ is defined as below:
\begin{equation}
D=6H^2+\frac{3}{2}(\frac{\dot{F}}{F})^2+6H\frac{\dot{F}}{F}-\frac{R}{2}+(\frac{k}{a})^2.
\\ \nonumber
\end{equation}
In the  modified gravity theories where the action is close to the
Einstein Hilbert (i.e $m\ll 1$),  the screened mass reduces to
\cite{Tsujikawa08}:
\begin{equation}
Q(k,z)\equiv \frac{1}{F}(1+\frac{\xi}{1-m}),
\end{equation}
where $\xi = mk^{2} a^{-2}R^{-1}$.

We note that in the definition of the gauge invariant density
contrast in Eq.(\ref{dc}), the extra term added to $\delta$ is in
the order of $v_Hv_p/c^2$, where $v_H$ is the velocity of Hubble
expansion at the size of the structure and $v_p$ is the
corresponding peculiar velocity. In order to have an estimation from
this term, we take Harisson-Zeldovich spectrum for the density
contrast. Then the density contrast of a structure changes as
$\delta_k = \delta_{enter} k^2/H^2$, where $\delta_{enter} =
10^{-5}$ is the density contrast of the structure with the horizon
size at the present time. On the other hand the peculiar velocity in
the linear regime is in the order of $v_k = H_0 \delta_k/k $.
Substituting $\delta_k$ from the Harisson-Zeldovich spectrum, the
peculiar velocity obtained as $v_k = k/H_0 \delta_{enter}$, or $v_k
v_H \simeq 10^{-5}$. For sub-horizon scales where $\delta_N \gg
10^{-5}$, we can neglect the second term in equation (\ref{dc}). The
results within this approximation is identical to the case where we
choose comoving gauge in the analysis. In this gauge, the peculiar
velocity is set to zero \cite{Koivisto2006}.

Another indication of the modified gravity is that, the right hand
side of the Eq. (\ref{EqShearGr}) in the late time universe is
non-zero. Using the projection operator of $\hat{k}_i\hat{k}^j - 1/3
\delta_i^j$ on the perturbed field equation of (\ref{pfe}) results
in \cite{Pogosian2008,Tsujikawa08,Koivisto2006}:
\begin{equation}
\zeta= \Phi+\Psi=-\frac{\delta F}{F}, \label{ps}
\end{equation}
where $\delta F$ represents the perturbation of the action due to
the perturbation of the metric which can be rewritten as $\delta F=
F_{,R}\delta R \equiv{dF}/{dR}~\delta R$. We note that for any
deviation of the gravity law from the Einstein-Hilbert action in Eq.
(\ref{ps}), $\Phi+\Psi \neq 0$ and the screened mass function is not
equal to one. This may affect on the growth of the large scale
structures and consequently modify the expected power spectrum of
the structures.

 The deviation of the Ricci scalar $\delta R$ due to the perturbation of the
energy-momentum tensor is given by the trace of field equation
(\ref{EqPalTrace}) as below:
\begin{equation}
\delta R= \frac{8\pi G \delta T}{RF_{,R}-F}, \label{dr}
\end{equation}
Where $\delta T$ is  trace of perturbed energy-momentum tensor and
in the case of matter dominated era is given by $\delta
T=-\delta\rho$. We substitute (\ref{dr}) in (\ref{ps}) and change
the derivatives from the Ricci scalar to redshift, so the $\zeta_k$
in Fourier mode is given by:
\begin{equation} \label{EqZeta}
\zeta_k\equiv\Phi+\Psi=\frac{3\Omega_{m}\delta_{m}(k,z)\frac{dF}{dz}(\frac{d\tilde{R}}{dz})^{-1}}{
F[\tilde{R}\frac{dF}{dz}(\frac{d\tilde{R}}{dz})^{-1}-F]},
\end{equation}
where $\Omega_{m}$ is the matter density parameter in redshift of
$z$, $\tilde{R}={R}/{H_{0}^2}$ and $\delta_{m}(k,z)$ is the matter
density contrast which depends on wavenumber and redshift.

Now we can define the gravitational slip parameter in terms of MG
free parameter $F$. Using Eqs. (\ref{gamma}) and (\ref{ps}), the
slip parameter can be written as:
\begin{equation}
\gamma(k,z)=1+\frac{F_{,R}\delta R}{\Psi F}, \label{gam1}
\end{equation}
On the other hand we want to write the gravitational slip parameter
independent of the matter density contrast. The combination of
Eqs.(\ref{EqScreenedMass}), (\ref{dr}) and (\ref{gam1}) results in:
\begin{equation}{\label{slip}}
\gamma(k,z)=[1-\frac{2\tilde{k}^{2}(1+z)^2\frac{dF}{dz}(\frac{d\tilde{R}}{dz})^{-1}}{
FQ(k,z)(\tilde{R}\frac{dF}{dz}(\frac{d\tilde{R}}{dz})^{-1}-F)}]^{-1}.
\end{equation}
We can also rewrite the gravitational slip parameter in terms of
deviation parameter as below:
\begin{equation}\label{slipf}
\gamma=[1-\frac{2m(1+z)^2{\tilde{k}}^2}{\tilde{R}FQ(m-1)}]^{-1}.
\end{equation}
In the following sections we will apply the gravitational slip
parameter in the Integrated Sachs-Wolfe effect (ISW) and study the
deviation of the CMB power spectrum from the dark energy models.

In order to find the evolution of the matter density contrast in MG,
We start with the conservation of energy momentum tensor in Eqs.
(\ref{Eqdc}) and (\ref{EqEul}). For the matter component of the
perturbation $\omega=0$ and for the scales larger than Jeans length
$\delta p=0$. The energy-momentum conservation equations reduce to:

\begin{equation}\label{EqdcMatter}
\dot{\delta}=-\frac{1}{a}\theta-3\dot{\Phi},
\end{equation}
\begin{equation}\label{EqEulMatter}
\dot{\theta}=-H\theta+\frac{k^2}{a}\Psi.
\end{equation}
Now in order to find the evolution of the density contrast, we take
the time derivative from Eq.(\ref{EqdcMatter}) and eliminate
$\dot\theta$ by substituting it from equation (\ref{EqEulMatter}).
The density contrast is obtained in terms of the potentials $\Psi$ ,
$\Phi$ as below:
\begin{equation}
\ddot{\delta}+2H\dot{\delta}+\frac{k^2}{a^2}\Psi=-6H\dot{\Phi}-3\ddot{\Phi}
\label{dfd}
\end{equation}
In order to determine the evolution of density contrast, we apply
the modified Poisson equation, Eq.(\ref{EqScreenedMass}) and the
gravitational slip parameter, Eq.(\ref{gamma}) in Eq.(\ref{dfd}). We
find the differential equation as below:
\begin{eqnarray} \label{DiffDc}
&[1&+3\frac{\Gamma}{ak^2}Q]\ddot{\delta}_{m}+[2H+6\frac{\Gamma}{ak^2}\dot{Q}]\dot\delta_{m}
\\ \nonumber &+&[-\gamma^{-1}\Gamma
a^{-3}Q-3\frac{\Gamma}{ak^2}(H^2Q+\dot{H}Q-\ddot{Q})]\delta_{m}=0,
\end{eqnarray}
where $\Gamma=4\pi G \rho^{0}_{m}$.  Eq.(\ref{DiffDc}) is the
general form of the differential equation for the evolution of
matter density. In the limit of quasi-static approximation
$\dot{\Phi}\approx 0$ with the assumption of studying the scales
deep inside the horizon, where $\frac{\Gamma}{k^2}\sim
(\frac{H}{k})^2\rightarrow 0$, Eq.(\ref{DiffDc}) reduces to the
equation defined as \cite{Amendola2010}:
\begin{equation}
\ddot{\delta}_{m}+2H\dot{\delta}_{m}-4 \pi
G\rho_{m}\delta_{m}\gamma^{-1}Q(k,z)=0.
\end{equation}
For the case of GR, we will have $\gamma=Q=1$ and we recover the
standard equation for the evolution of density contrast
{\cite{Tsujikawa08}}.

In the following sections we will solve this differential equation
for a suitable modified gravity, reconstructed from the SNIa data.

\subsection{ $f(R)$ Modified gravity theories in Metric Formalism}

\label{subsection3-2} As we mentioned earlier there are two
different approaches for the modified gravity models. It seems that
one of the possible tools to distinguish them is the evolution of
the large scale structures in these two formalism. Here in this
section we study the the growth of structures in the metric
formalism and obtain the evolution of density contrast and the
screened mass function. Further investigation in metric formalism
for the non-linear regime of the structure formation will be done in
our future work.

The action of the modified gravity in metric formalism is given by:
\begin{equation} \label{MetricAction}
S=\frac{1}{16\pi G}\int
d^4x\sqrt{-g}f(R)+S_{m}(g_{\mu\nu},\Psi_{m}),
\end{equation}
where the connections $\Gamma^{\alpha}_{\beta\gamma}$ are usual
connections defined in terms of the metric $g_{\mu\nu}$ . The field
equations obtained by varying the action (\ref{MetricAction}) with
respect to $g_{\mu\nu}$:
\begin{equation}\label{ActionMetric}
F(R)R_{\mu\nu}(g)-\frac{1}{2}f(Rg_{\mu\nu}-\nabla_{\mu}\nabla_{\nu}F(R)+g_{\mu\nu}\Box
F(R)=8\pi GT_{\mu\nu}
\end{equation}
and its trace is
\begin{equation}{\label{metricTrace}}
3\Box F(R)+F(R)R-2f=8\pi GT.
\end{equation}

The evolution of density contrast  will be derived from perturbed
Einstein equations in deep inside the Hubble radius is (for more
details see Appendix A):
\begin{equation}
-\frac{k^2}{a^2}\Phi+3H(H\Psi-\dot{\Phi})=\frac{1}{2F}\left[3H\dot{\delta
F}-\left( 3\dot{H}+3H^2-\frac{k^2}{a^2}\right)\delta
F-3H\dot{F}\Psi-3\dot{F}(H\Psi-\dot{\Phi})-\kappa\delta\rho)\right],
\end{equation}
\begin{equation}
\ddot{\delta F}+3H\dot{\delta F}+(\frac{k^2}{a^2}-\frac{R}{3})\delta
F=\frac{\kappa}{3}\delta\rho_{m}+\dot{F}(3H\Psi+\dot{\Psi}-3\dot{\Phi})+(2\ddot{F}+3H\dot{F})\Psi-\frac{1}{3}F\delta
R,
\end{equation}
\begin{equation}
\Psi+\Phi=-\frac{\delta F}{F}.
\end{equation}
In the quasi-static regime where $|\dot{F}|\ll|HF|$ and
$|\ddot{F}|\leq|H^2F|$, the dominate terms will be $k^2/a^2$ and
$\kappa\rho$ and with the assumption of small deviation from GR the
equation governing the evolution of matter density will reduce to
\cite{Amendola2010}:
\begin{equation}\label{EqMetricDensity}
\ddot{\delta}_{m}+2H\dot{\delta}_{m}-4\pi G\gamma_{met}^{-1}
Q_{met}\delta\rho_{m}\approx 0,
\end{equation}
where $\gamma_{met}$ is the gravitational slip parameter and
$Q_{met}$ is the screened mass  function, both defined in the metric
formalism. The screened mass function in the metric formalism is
given by:
\begin{equation}\label{Q-metric}
Q\equiv G_{eff}/G=
\left[\frac{4}{3}-\frac{1}{3}\frac{M^2a^2}{k^2+M^2a^2}\right],
\end{equation}
where $M^2\equiv\frac{F}{3}(\frac{\partial F}{\partial R})^{-1}$,
which in the case of $M\rightarrow\infty$  we will recover the GR
case $Q=1$. The equation above indicates that in large scales $k\ll
Ma$, we will recover $Q_{met}\rightarrow 1$, while in the small
scales $k\gg Ma$, the screened mass function converge to $4/3$.
Meanwhile by assumption of the same background dynamics, in the
Palatini formalism  the screen mass function from
Eq.(\ref{EqPalScreenedMass})  predicts a different scale dependance.
Consequently the structure formation observables will have different
values for metric and Palatini formalisms. It is crucial point to
indicate that for this comparison, we should also extract the
corresponding modified gravity action in the metric formalism. For
this task we refer to modified Friedmann Equations in metric
formalism obtained from field equation (\ref{ActionMetric}):
\begin{eqnarray}\label{metricFriedmann}
3FH^2=8\pi G\rho_{m}+\frac{RF-f}{2}-3H\dot{F}\\ \nonumber
-2F\dot{H}=8\pi G\rho_{m}+\ddot{F}-H\dot{F}
\end{eqnarray}
where dot represent derivative with respect to physical time which
can be reexpressed in terms of redshift as:
\begin{equation}\label{Fmetric}
F^{''}+\left[\frac{2}{1+z}+\frac{E^{'}}{E}\right]F^{'}-\frac{2F}{1+z}\frac{E^{'}}{E}=-\frac{3\Omega^{0}_{m}(1+z)}{E^2}.
\end{equation}
By substituting the reconstructed Hubble parameter
$E=\frac{H^2}{H^2_{0}}$ in Eq (\ref{Fmetric}) from the equivalent
sDE model and solving the differential equation with GR boundary
conditions in higher redshifts (i.e. $z\sim100$), we can find
$F=F(z)$ and eliminating the Ricci scalar derive the action
\cite{rahvar}. We will review the procedure of deriving the
equivalent action with a given dark energy model in Palatini
formalism in the following section.

\section{ Structure Formation: Modified gravity versus Dark energy}

\label{section4} In this section we first review the dynamical
equivalence of a modified gravity with a smooth dark energy model.
Knowing the dynamics of the universe from the observational data, we
extract the appropriate action for the modified gravity as described
in a recent work by Baghram and Rahvar \cite{Baghram09}. From the
action, we obtain the screened mass function and the gravitational
slip parameter. Then we calculate the evolution of the large scale
structures for the MG and compare it with that from the dark energy
equivalence, assuming the same initial condition for the density
contrast. Finally we compare the power spectrum of the structures as
a discriminating tool in these two models.

\subsection{Geometrical Equivalence of the Modified gravity theories with the
smooth Dark Energy models} In this part, we review the equivalence
of dark energy and modified gravity in the background dynamics of
the universe (i.e. $H=H(z)$) \cite{Baghram09}. In the standard GR,
from the Friedman equations the energy density is related to the
dynamics of the background as:
\begin{eqnarray}
H^{2}=\frac{\kappa}{3}\rho, \\ \nonumber
\frac{\ddot{a}}{a}=-\frac{\kappa}{6}[\rho+3p],
\end{eqnarray}
where $\rho$ and $p$ are the overall energy density and the pressure
of the cosmic fluid. The effective equation of the state for the
cosmic fluid can be written as:
\begin{equation} \label{EqOmegaeff}
\omega_{eff}=-1-\frac{2}{3}\frac{\dot{H}}{H^2}.
\end{equation}
Separating the contribution of the dark matter and dark energy in
the matter budget of the universe, the equation of state of the dark
energy is related to the effective equation of state as follows:

\begin{equation}\label{EqOmegaDE}
\omega_{DE}(z)=\frac{E^2(z)\omega_{eff}}{E^2(z)-{\Omega_{m}^{0}(1+z)^3}},
\end{equation}
where $\Omega_{m}^{0}$ is the matter density parameter at the
present time. For a given modified gravity model, we can substitute
the Hubble parameter in equations (\ref{EqOmegaeff}) and
(\ref{EqOmegaDE}) to obtain the equivalent equation of state of a
dark energy model. It should be noted that the geometrical
equivalence between the two models is a one-to-one relation.

\FIGURE{\epsfig{file=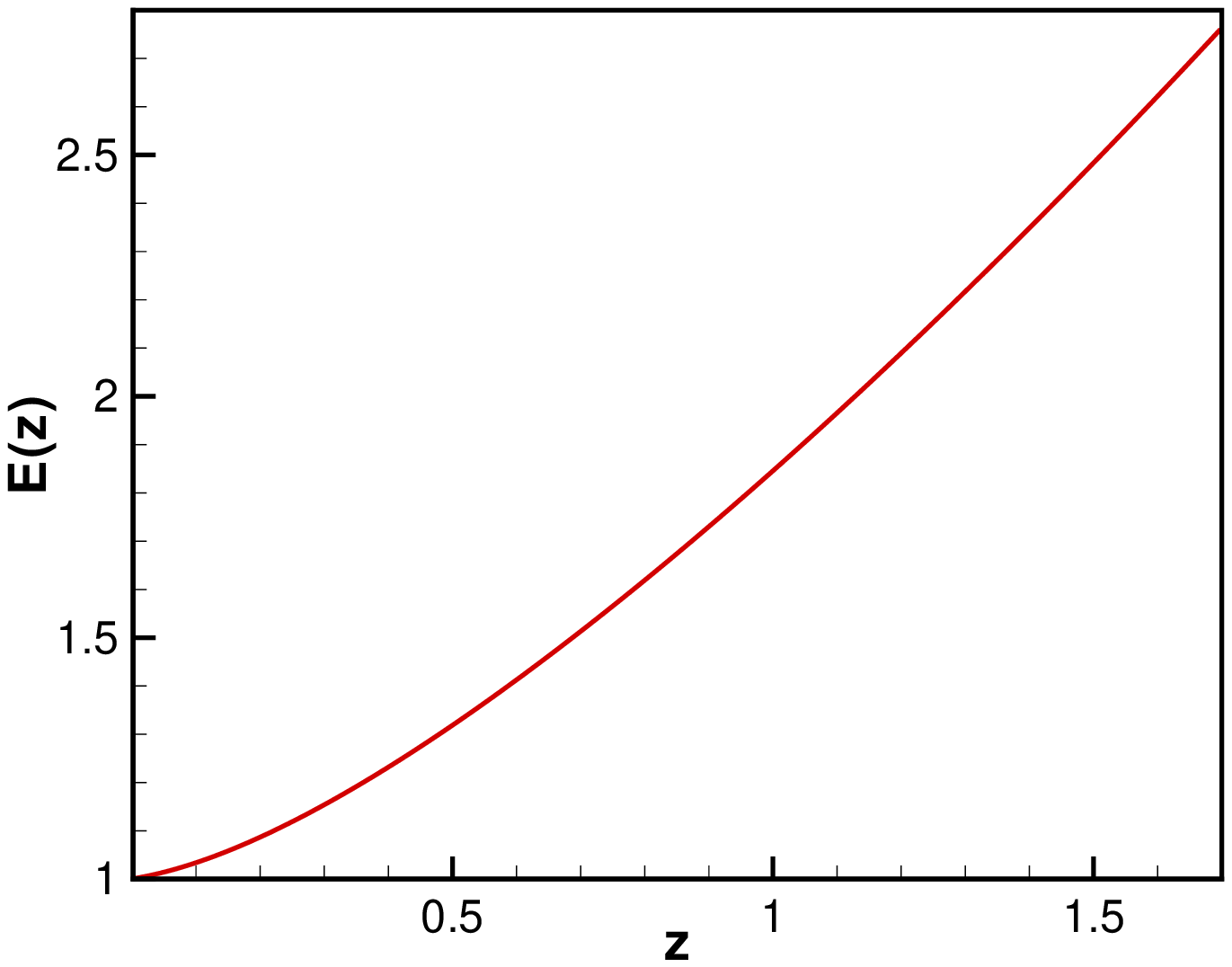,width=300pt}\caption{ The
dimensionless Hubble parameter, $E(z)=H(z)/H_{0}$ versus redshift
derived from solving the Eq.(\ref{EqOmegaDE}) for a dark energy
model with the equation of state defined in (\ref{EqCPL}).}
\label{fig1}}

\FIGURE{ \epsfig{file=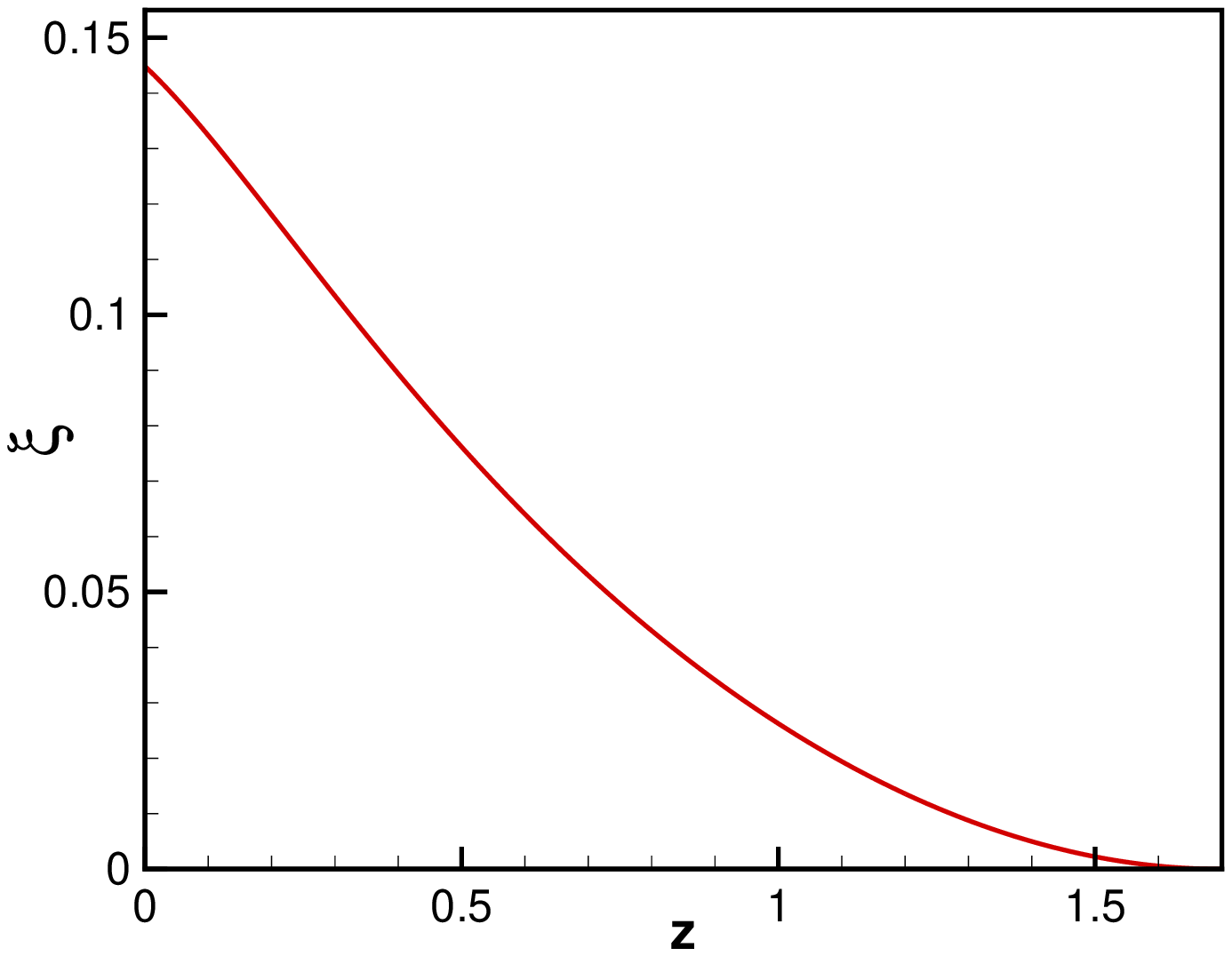,width=300pt}
\caption{The deviation parameter $\xi={df}/{dR}-1$ versus redshift
derived from solving the differential equation (\ref{difftau}). Here
the modified gravity is obtained from the dynamical equivalance of a
dark energy with an equation of state given in (\ref{EqCPL}).}
\label{fig2} }

\FIGURE{\epsfig{file=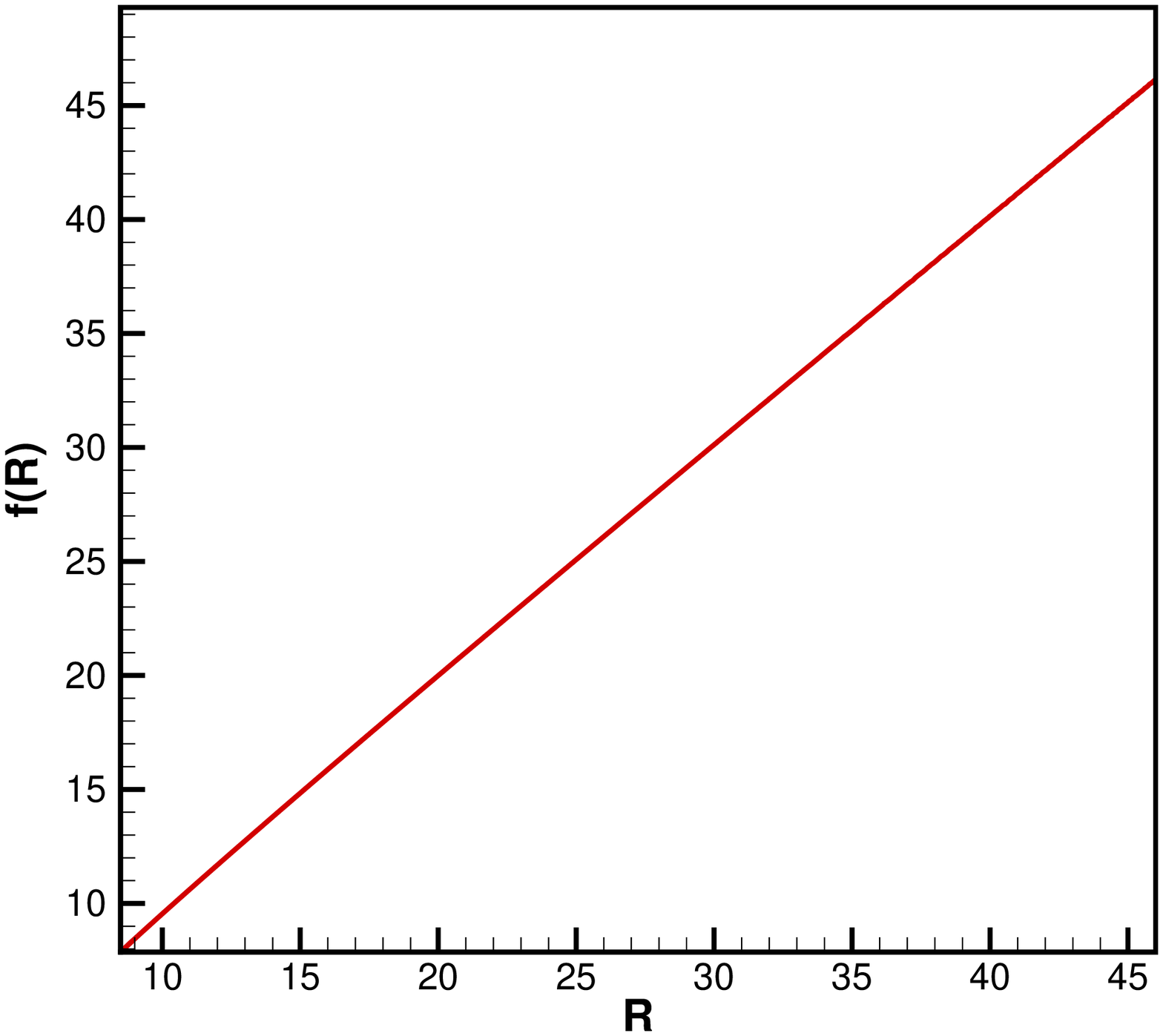,width=300pt}\caption{ The modified
gravity action $f(R)$, which is obtained from integration of
deviation parameter $\xi$} \label{figfR}}

On the other hand, for a given equation of state of a dark energy
model, $\omega_{DE}(z)$, one can extract the dynamics of the
universe, $E(z)$. Knowing the dynamics of the universe, from
Eq.(\ref{EqPalHubble}), the Hubble parameter is given in terms of
the derivatives of the action and the Ricci scalar. The Ricci scalar
in Palatini formalism is given by:
\begin{equation} \label{Ricpal}
R=R(g)+\frac{3}{2}\frac{\nabla_{\alpha}F\nabla^{\alpha}F}{F^{2}}-\frac{3\nabla_{\alpha}\nabla^{\alpha}F}{F}æ,
\end{equation}
in which $R(g)$ is the Ricci scalar in terms of the metric and
$R(g)=6\dot{H}+12H^{2}$. For the FRW metric, the Ricci scalar in the
Palatini formalism is given by \cite{Sotiriou2008}:
\begin{eqnarray} \label{Ricci1}
&R&=-6HH'(1+z)+12H^{2}-\frac{3}{2}H^2(1+z)^2(\frac{F'}{F})^2\nonumber
\\ &+&3(1+z)^2HH'\frac{F'}{F}-6H^2(1+z)\frac{F'}{F}\nonumber \\
&+&3H^2(1+z)^2\frac{F''}{F}.
\end{eqnarray}
Now we substitute the Ricci scalar in Eq.(\ref{EqPalHubble}) and
with some algebraic manipulation we obtain a differential equation
for the evolution of $F$ in terms of redshift:
\begin{eqnarray}
F'' &-& \frac32\frac{F'^2}{F} + F'(\frac{H'}{H} + \frac{2}{1+z})
\nonumber \\
&-&\frac{2H'}{H(1+z)}F + \frac{\kappa}{H^2(1+z)^2}(\rho+p) = 0,
\label{difftau}
\end{eqnarray}
where prime represents the derivative in terms of the redshift.
Assuming $p=0$ and $\rho = \rho_0(1+z)^3$, we eliminate the density
in favor of the Hubble constant, using $\kappa\rho_0 =
3H_0^2\Omega_{m}^0$. Finally we can solve the differential equation
for a given Hubble parameter and obtain $F$ in terms of redshift.

Here we choose a smooth dark energy ansatz with the equation of
state of \cite{CPL}:
\begin{equation} \label{EqCPL}
\omega=\omega_{0}+\omega_{a}\frac{z}{1+z},
\end{equation}
where $\omega_{0}$ and $\omega_{a}$ are the free parameters of the
model. In the case of $\omega_{0}=-1$ and $\omega_{a}=0$ we recover
the $\Lambda$CDM model. Here in our simulation we choose
$\omega_{0}=-1.2$, $\omega_{a}=1.5$ and use Eqs. (\ref{EqOmegaeff})
and (\ref{EqOmegaDE}) to extract the corresponding Hubble parameter
for this equation of state as shown in Fig.(\ref{fig1}). On the
other hand substituting the Hubble parameter in Eq. (\ref{difftau})
results in $F$ in terms of redshift. We set the boundary conditions
in redshift of $z\sim100$, where we anticipate to recover the
general relativity condition (i.e. $F(z=100)=1$ and
$F^{'}(z=100)=0$). Applying the Hubble parameter in
Eq.({\ref{Ricci1}}) we get Ricci scalar in terms of the redshift. On
the other hand we substitute the Hubble parameter in (\ref{difftau})
and get $F$ in terms of the redshift. We plot the parameter of the
action deviation, $\xi\equiv F-1$ as a function of $z$ in Fig.
(\ref{fig2}). Eliminating the redshift between the $F$ and $R$
results in the derivative of the action in terms of the Ricci
scalar, $F = F(R)$. One step further we do integration and find the
modified gravity action $f(R)$ in terms of $R$,
as shown in Fig.(\ref{figfR}). A numerical  fit to this function is:
\begin{equation}\label{f}
f(R)\sim R-a \textrm{e}^{-\frac{R}{R_{0}}},
\end{equation}
where $a\simeq0.1$ and $R_{0}\sim 8.48 H_{0}^2$ is the Ricci scalar
at the present time. It is obvious from the form of Eq.(\ref{f})
that for large scalar Ricci we will have the Einstein-Hilbert
action.

It is worth to mention that for a smooth potentials of the scalar
fields in the quintessence models, Bassett et al. (2002) and
Corasaniti et al. (2003) \cite{Bassett} show that the equivalence
dark energy model with this class of the quintessence models can be
identify with two free parameters of $w_0$ and the derivation of the
$w$ at the present time (i.e. $w(z) = w_0 + \frac{w_az}{1+z}$). We
did our analysis based on this class of the dark energy models and
didn't test in the cases that $w(z)$ has dramatic variation with to
redshift.

\subsection{Reconstruction of the Action in the Modified Gravity}

\FIGURE{ \epsfig{file=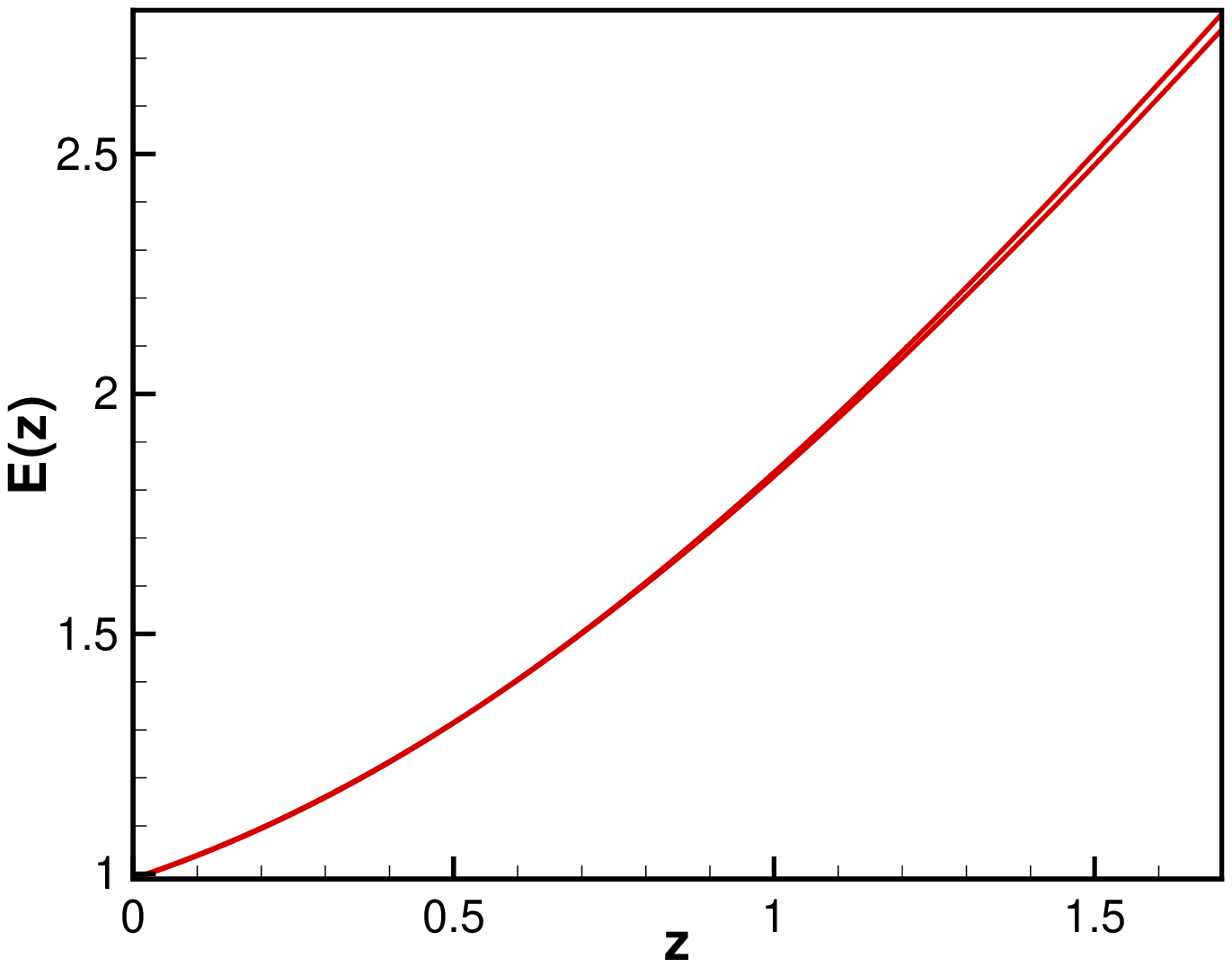,width=300pt}
\caption{The reconstructed Hubble parameter with 1$\sigma$
confidence level obtained from the reconstruction of 100 ensemble of
SNAP data.} \label{fig3}}

\FIGURE{\epsfig{file=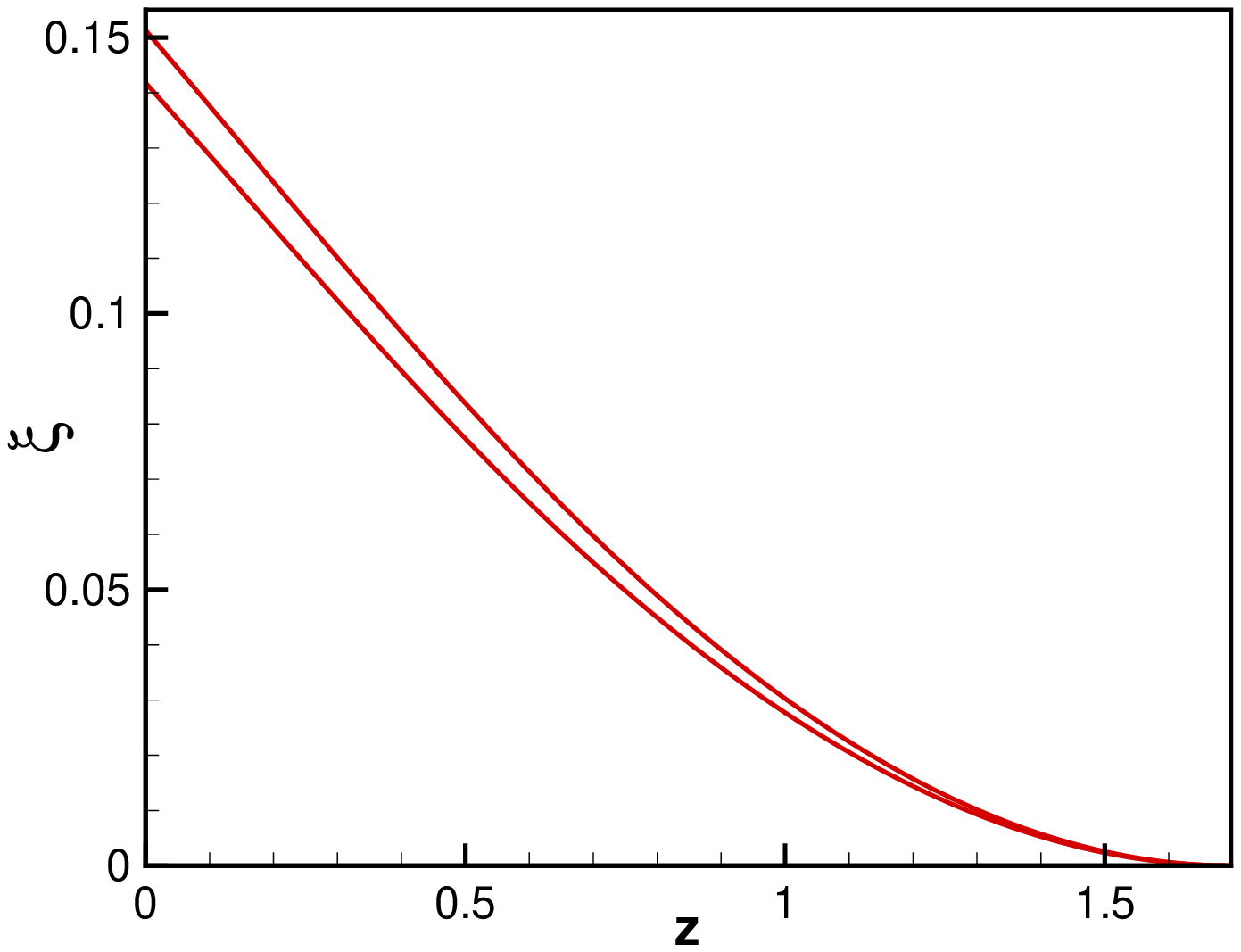,width=300pt}  \caption{The
reconstructed deviation parameter $\xi={df}/{dR}-1$ with 1$\sigma$
confidence level obtained from the synthetic SNAP data. The
uncertainly in $\xi$ indicated with the solid lines, represents
different realizations of SNIa data in an ensemble of 100 data set.}
\label{fig4}}

Our approach for reconstruction of the action is based on the
measurement of the distance modulus of the SNIa data. From the SNIa
data we can make a continues function for the distance modulus as a
function of redshift (i.e. $\mu(z)$), using the smoothing method \cite{Cappo2005,Starobinsky98,Alderin2004}
as described in \cite{Shafieloo2006}. The smoothing method suffers from
the low statistics of the SNIa data where it has been shown in
\cite{Baghram09} that at least $\sim 1500$ SNIa data with the
quality of Union sample data set \cite{kowalski08} is essential to
generate a reliable distance modulus.

Here we repeat the same procedure with 1998 synthetic SNIa data with
the quality predicted by SNAP project in the redshift range of
$0.1<z<1.7$. We also add 25 SNIa data from Nearby Supernova factory
project in the redshift of $z<0.1$ to have a full range coverage of
SNIa data \cite{Alderin2004,Shafieloo2006}. Now we use the
hypothetical cosmological model as described in Eq. (\ref{EqCPL})
for generating the corresponding luminosity distance for the SNIa
data with a systematic error of $\sigma_{sys}=0.07$ in SNAP.

After generating the synthetic distance modulus, we obtain the best
curve resulting from the smoothing of the simulated data. For the
smoothing discrete data, we first take an initial guess model for
the distance modulus based on $\Lambda CDM$ model. Then we smooth
the subtraction of the distance modulus of simulated data from the
guess model using a Gaussian function as below:
\begin{equation}
\mu^{d}(z)=N(z)\sum_i[\mu^{g}(z)-\mu^{sim}(z_{i})]\exp{\frac{-(z-z_{i})^{2}}{2\Delta^{2}}},
\label{smooth1}
\end{equation}
where
\begin{equation}
N(z)^{-1}=\sum_i \exp[\frac{-(z-z_{i})^2}{2\Delta^2}],
\end{equation}
$z_{i}$ represents the redshift of each SNIa in the simulated SNAP
sample. The sum term is considered for all 2023 SNAP simulated data
sample, $\mu^{sim}(z_{i})$ is the simulated distance modulus and
$\mu^{g}(z)$ is a continues function generated by the guess model.
$N(z)$ is a normalization factor, $\Delta$ is a suitable redshift
window function and $\mu^{d}(z)$ is a smoothed continues function
for the residual of luminosity distance in terms of redshift.
Finally we add the smoothed distance modulus to the guess model to
generate the new smooth function for the distance modulus:
\begin{equation} \label{smooth}
\mu^{s}(z)=\mu^{g}(z)+\mu^{d}(z).
\end{equation}
We repeat this procedure using $\mu^s(z)$ as the new guess model. It
can be shown that after a finite time of irritations, the $\chi^2$
of the smoothed function with respect to observed data will converge
to a fixed value\cite{Shafieloo2006}. Now we have a continues
distance modulus function in which the Hubble parameter in terms of
redshift can be obtained by:
\begin{equation}
H(z)=[\frac{d}{dz}(\frac{d_{L}(z)}{1+z})]^{-1}. \label{rec}
\end{equation}
In order to calculate the uncertainty of this procedure, we generate
an ensemble of the SNIa data and extract the corresponding Hubble
parameters. Fig.(\ref{fig3}) shows the reconstructed Hubble
parameter with 1$\sigma$ level of confidence.

Now we can extract the modified gravity action in Palatini formalism
through the method of inverse problem \cite{rahvar} in the Palatini
formalism as discussed in the last part \cite{Baghram09}. Applying
the Hubble parameter from Eq.(\ref{rec}) in Eq.(\ref{difftau}), we
obtain $\xi$ in terms of redshift as plotted in Fig. (\ref{fig4})
with 1$\sigma$ level of confidence. Numerical integration of $F(R)$
provides the action, $f(R)$.

\subsection{Reconstruction of structure formation indicators and  matter density Power Spectrum}

In this section by applying the reconstructed action, we obtain the
density contrast and corresponding power spectrum of the matter and
compare it with that from the dark energy model. We start the
procedure by reconstructing the screened mass function by
substituting the numerical value for the first derivative of the
action in Eq.(\ref{EqPalScreenedMass}).
The corresponding screened mass function for this action is plotted
in Fig.(\ref{fig5}). At redshifts above $z \gtrsim1.5$, the screened
mass function approaches to unity in which the GR structure
formation is recovered. It should be noted that the screened mass
function does not  depend only on the chosen MG action but also it
depends on the wavenumber of the structures. One step further to
indicate the probable different predictions of metric and Palatini
formalism in matter density power spectrum we also plot the screened
mass function in metric formalism $Q^{met}(k,z)$ in
Fig.(\ref{fig-metric}) by applying the reconstructed dynamics of
universe and MG action in Eq.(\ref{Q-metric}).

\FIGURE{\epsfig{file=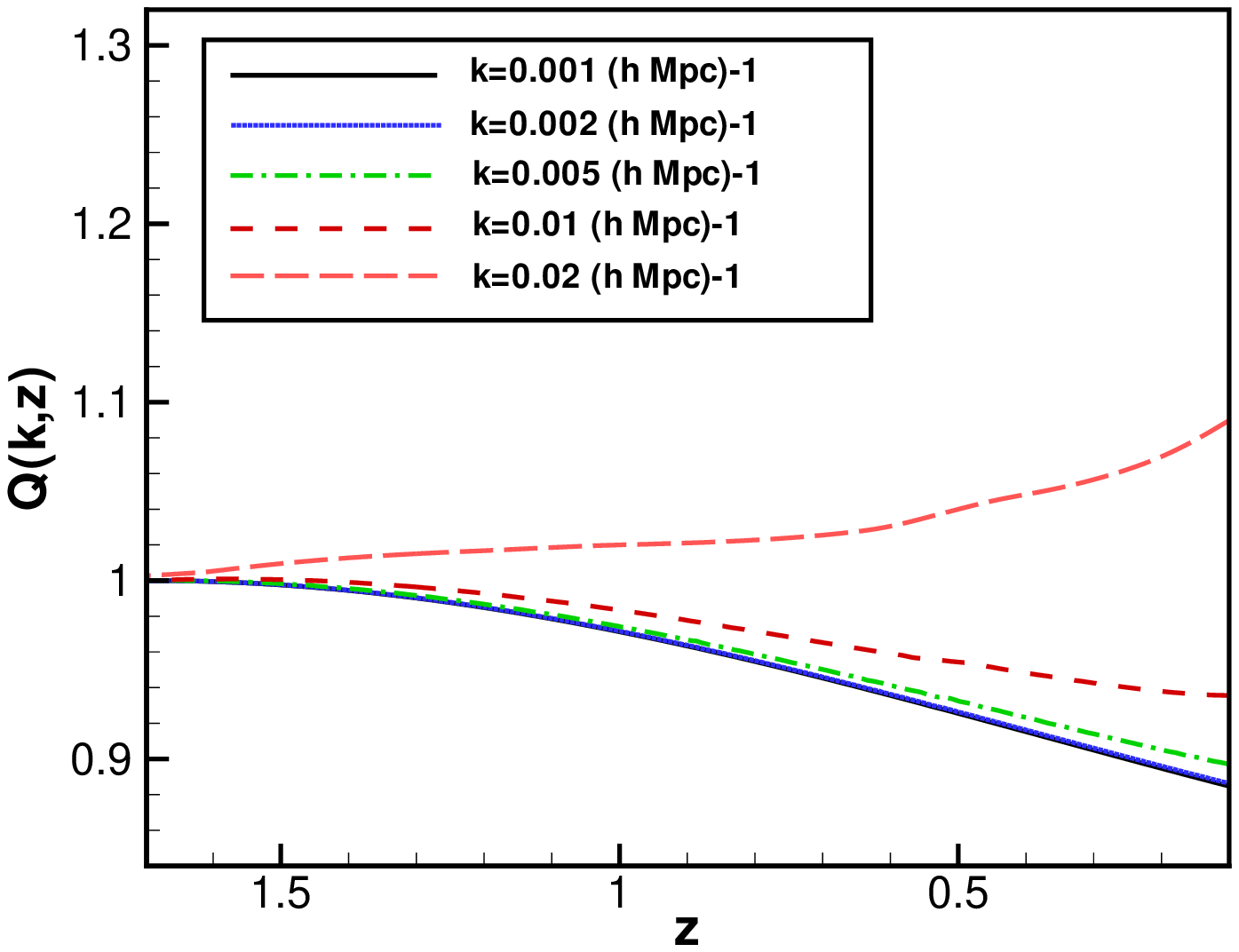,width=300pt}  \caption{Screened mass
function in Palatini formalism in terms of the redshift and
wavenumbers of the structures for the action equivalent to the dark
energy with the equation of state in Eq. (\ref{EqCPL}).
\label{fig5}}}

\FIGURE{\epsfig{file=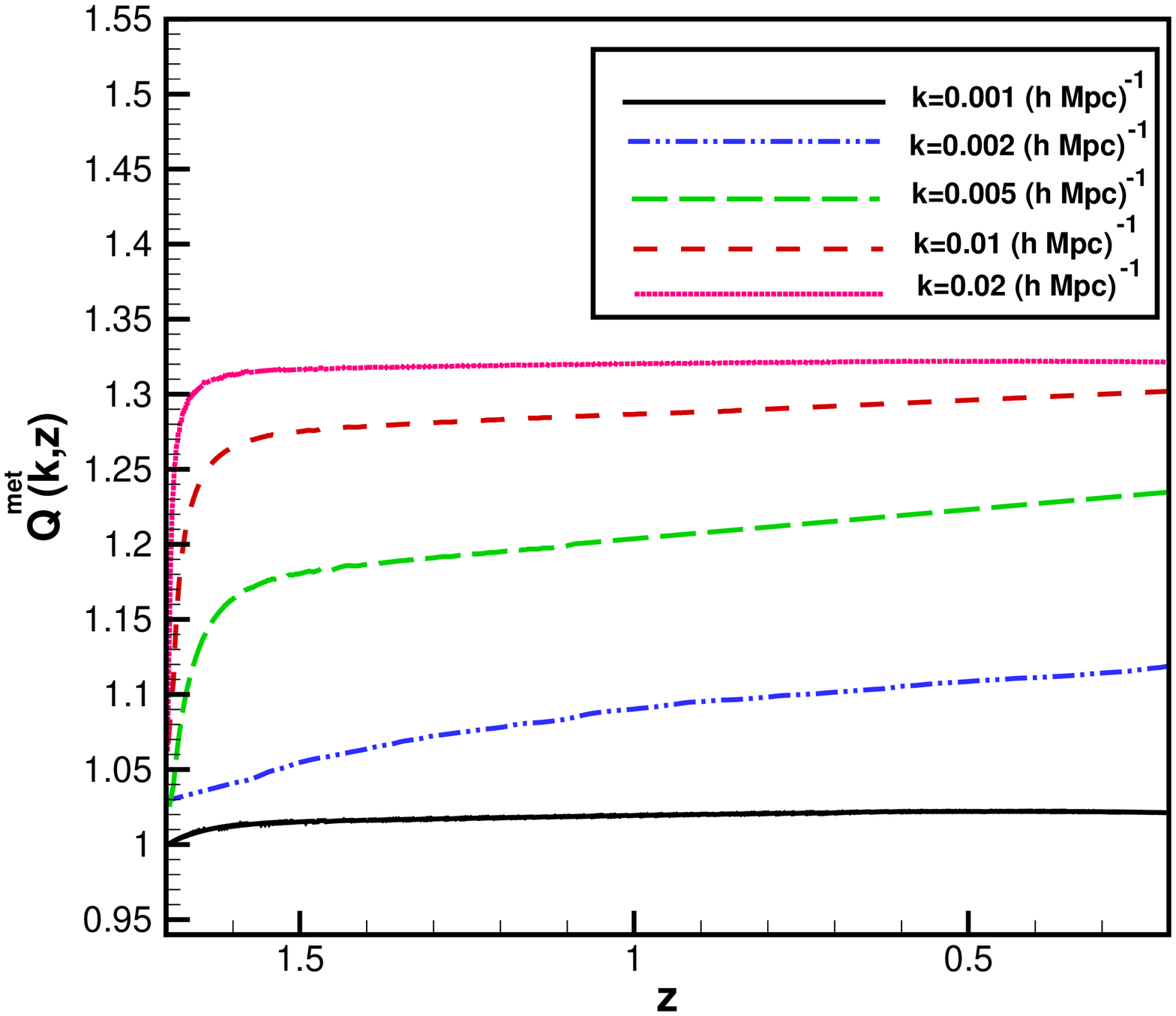,width=300pt}\caption{ Screened
mass function in metric formalism in terms of redshift and
wavenumbers of the structures for the action equivalent to the dark
energy with the equation of state in Eq. (\ref{EqCPL}).
}{\label{fig-metric}}}

The comparison of two Figures (\ref{fig5}) and (\ref{fig-metric})
shows the different scale dependence of screened mass function in
both formalisms and consequently the different effects on the
modification to the Poisson equations. A crucial point to indicate
is that this difference arise from the structure formation effects.
This is because  the background dynamics $H=H(z)$, is the same for
two formalisms obtained independently from SNIa data. The difference
comes from the definition of Ricci scalar $R(g)=6\dot{H}+12H^2$  and
modified Friedmann Eq.(\ref{metricFriedmann}) in metric formalism
and also the screened mass function defined in Eq.(\ref{Q-metric}).

In Palatini formalism, in  small wavenumbers, correspond to the
larger structures, $Q(k,z)$ descends for the later times, means
weakening of the gravitational strength in the Poisson equation. On
the other hand, in the smaller structures $Q$ is larger than one
(see. Fig.\ref{fig5}) which makes a faster growth of the structures
compared to the GR. The dependence of $Q$ to the wavenumber is
similar to the dispersion relation in optics in which different
length structures' gravitational response is different
\cite{Tsu2009}. The deviation of $Q$ from unity and also its
dependence on $k$ affects the growth of the density contrast and
subsequently the power spectrum of the structures. In Palatini
formalism the deviation from $Q=1$ is large at the larger scales, in
contrast in metric formalism we always have a faster growth in
structures and the the deviation is larger in small scales. These
differences can put their fingerprint in LSS observations. In what
follows we will solve the differential equation governing the
density contrast for various wavenumbers and finally obtain the
corresponding power spectrum of the structures.

\FIGURE{\epsfig{file=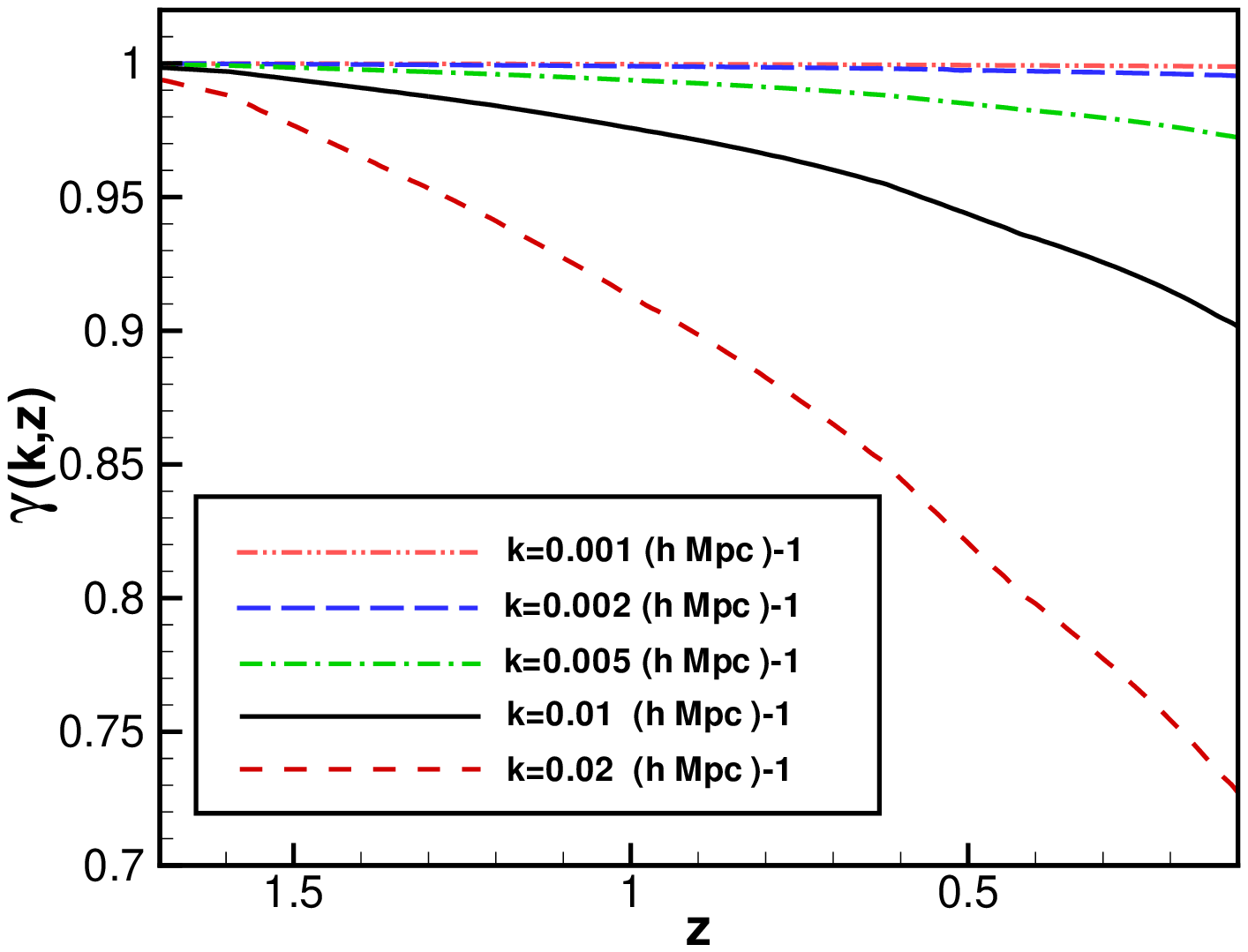,width=300pt}  \caption{The
gravitational slip parameter is plotted versus redshift for
different wavenumbers of the structures.} \label{fig6}}

In order to solve Eq.(\ref{DiffDc}) we need to obtain the
gravitational slip parameter. Substituting the action in
Eq.(\ref{slipf}), we reconstruct the gravitational slip parameter as
shown in Fig.(\ref{fig6}). The gravitational slip parameter is equal
to one for higher redshifts and GR is recovered. For the later times
at lower redshifts $\gamma$ deviates from GR. This deviation is
larger for the small structures. The variation of $Q$ and $\gamma$
with respect to GR indicates that the power spectrum of the
structures should deviate in comparison with the DE models. It is
worth to mention that the cosmic shear statistics as well as ISW
effect is sensitive to the gravitational slip parameter and provides
a promising tool for distinguishing between the models.

For calculating the evolution of the density contrast we reexpress
Eq.({\ref{DiffDc}}) in terms of redshift as below:
\begin{equation} \label{Eqdcredshift}
\delta_k^{''}+A(k,z)\delta_k^{'}+B(k,z)\delta_k=0,
\end{equation}
where the coefficients A and B are defined as:
\begin{eqnarray} \label{A}
&A&=\frac{E^{'}}{E}+\frac{1}{1+z}-\alpha(\tilde{k},z)(\frac{2}{1+z}-\frac{9(1+z)\Omega^{0}_{m}}{\tilde{k}^2}Q^{'}),\\
\nonumber
&B&=\alpha(\tilde{k},z)[-\frac{3}{2}\gamma^{-1}\Omega^{0}_{m}\frac{Q(1+z)}{E^2}-\frac{9\Omega^{0}_{m}}{2\tilde{k}^2}
(\frac{Q}{1+z}-Q\frac{E^{'}}{E}\\
&-&(1+z)Q^{'}\frac{E^{'}}{E}-Q^{'}-(1+z)Q^{''})], \label{B}
\end{eqnarray}
and
$\alpha(\tilde{k},z)\equiv{2\tilde{k}^2}/[{2\tilde{k}^2+9\Omega_{m}^{0}Q(1+z)}]$,
where ${\tilde{k}\equiv {k}/{H_{0}}}$. For the case that
$Q^{'}=Q^{''}\sim0$ and $\tilde{k}\gg 9\Omega_{m}^{0}Q(1+z)$ we will
recover the general form of density contrast evolution as:
\begin{equation}\label{EqPAlatiniDensityRedshift}
\delta^{''}+\left[\frac{E^{'}}{E}-\frac{1}{1+z}\right]\delta^{'}-\frac{3}{2}\frac{\Omega^{0}_{m}(1+z)}{E^2}\gamma^{-1}Q\delta=0.
\end{equation}
In this work within the Palatini formalism we solve the general form
of Eq.(\ref{Eqdcredshift}). The result for the reconstructed
modified gravity is plotted in Fig.(\ref{fig7}) for the various wave
numbers. Here we take the initial condition of the density contrast
from the Harrison-Zeldovich spectrum before the turn over point of
the power spectrum around $k\simeq 0.02$.
\FIGURE{\epsfig{file=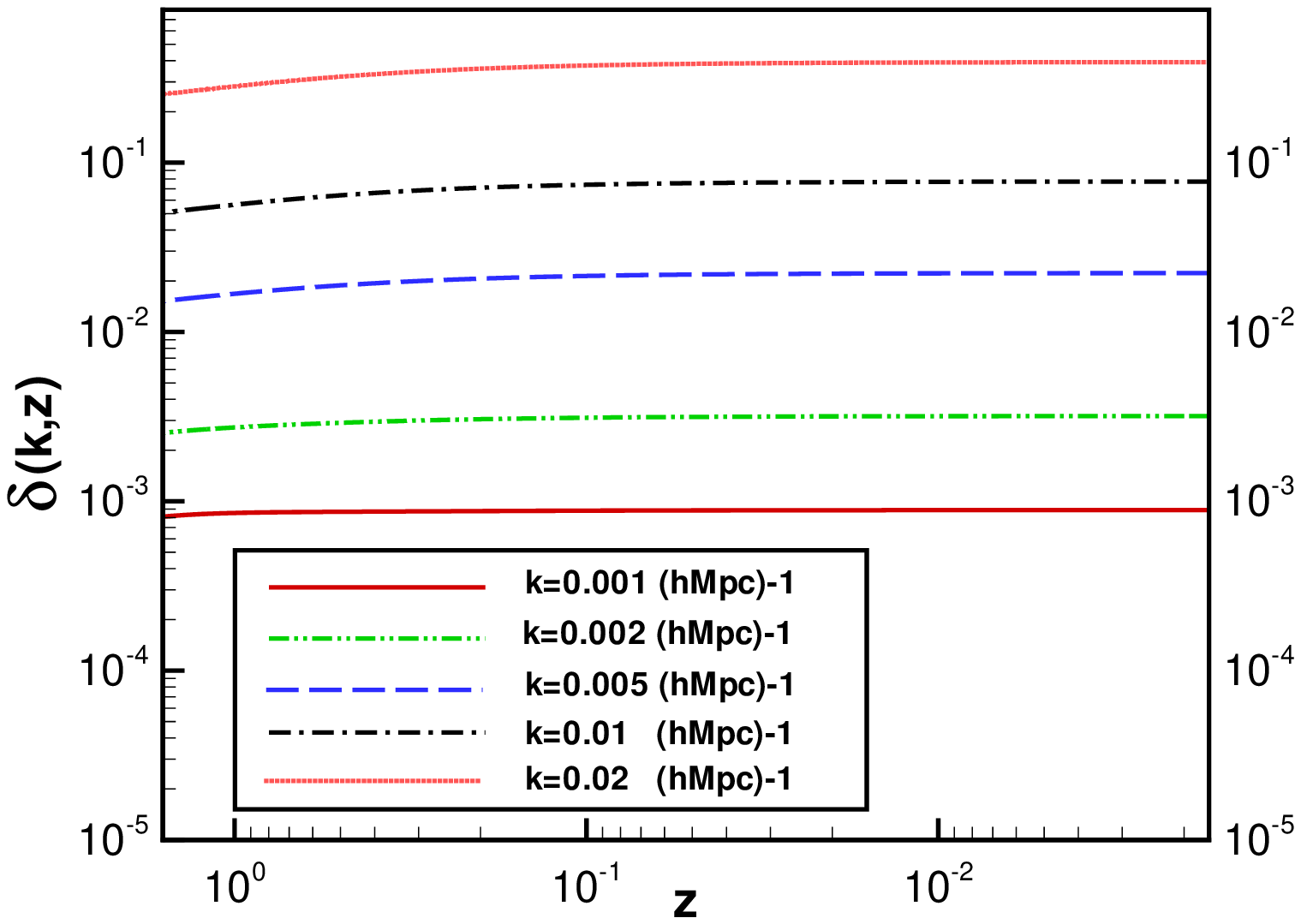,width=300pt}  \caption{The
matter density contrast in Palatini formalism is plotted versus
redshift for different wavenumbers.}\label{fig7}}

\FIGURE{\epsfig{file=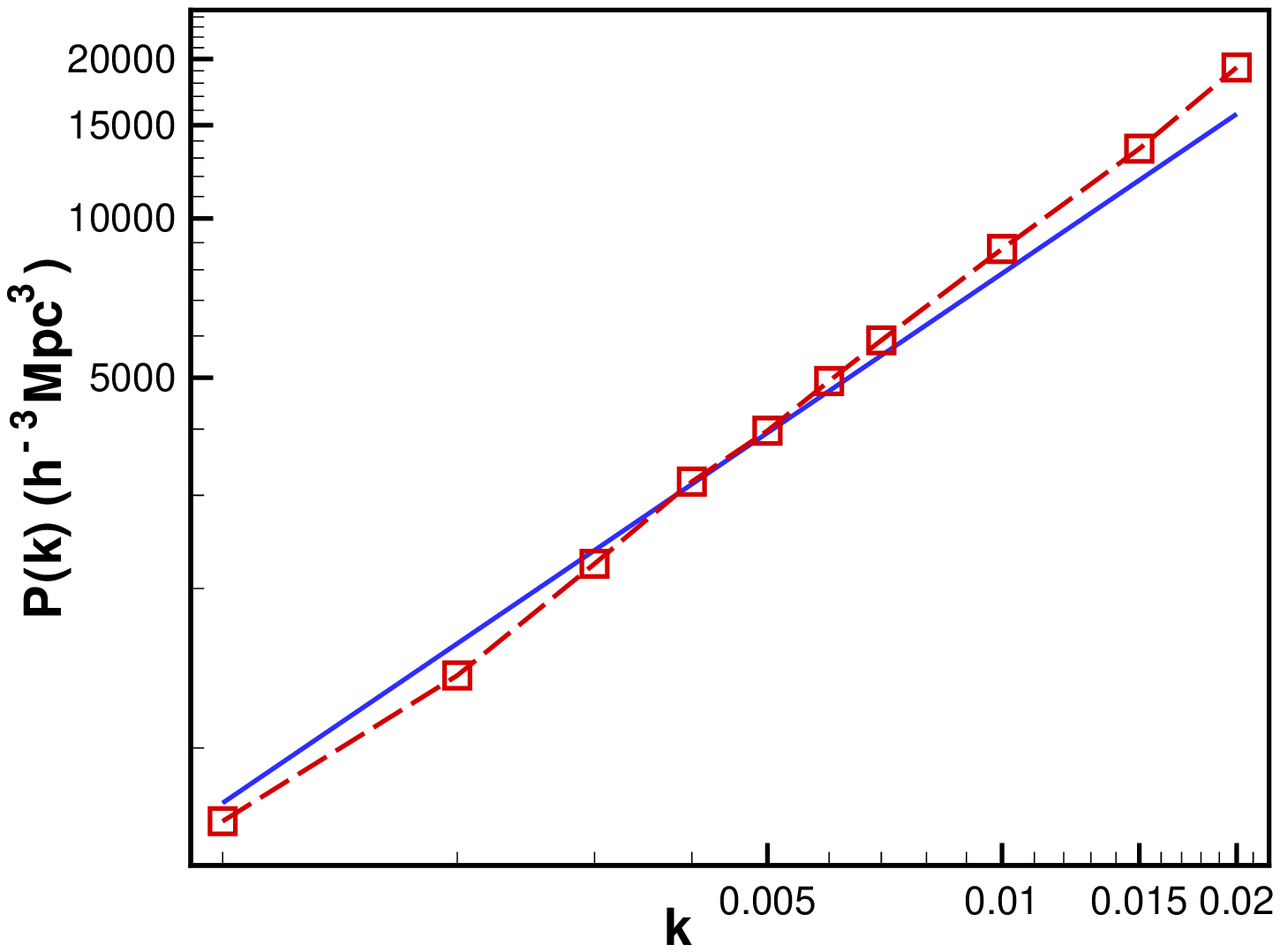,width=300pt}  \caption{The power
spectrum of MG theory in terms of wavenumber(dashed line) is plotted
in comparison to Harrison-Zeldovic power spectrum (solid line). For
the MG model the slop of the power spectrum is $n_f = 1.21^{
+0.13}_{-0.19}$.} \label{fig8}}

By solving Eq.(\ref{Eqdcredshift}) we obtain the power spectrum of
the structures at the present time as plotted it in Fig.
(\ref{fig8}) where the slope is $n_f = 1.21^{ +0.13}_{-0.19}$. It
should be noted that at these large scales (i.e. $k<0.01$) we don't
have accurate data from the large scale structure observations
\cite{{tegmark}}. In order to compare the two models with nowadays
LSS survey data, we should study the nonlinear regime of power
spectrum which is not in the scope of this work. We note that the
transformation of the linear power spectrum to nonlinear regime by
analytical methods is strictly model dependent and it is not yet
well understood \cite{Smith2003}.

\FIGURE{ \epsfig{file=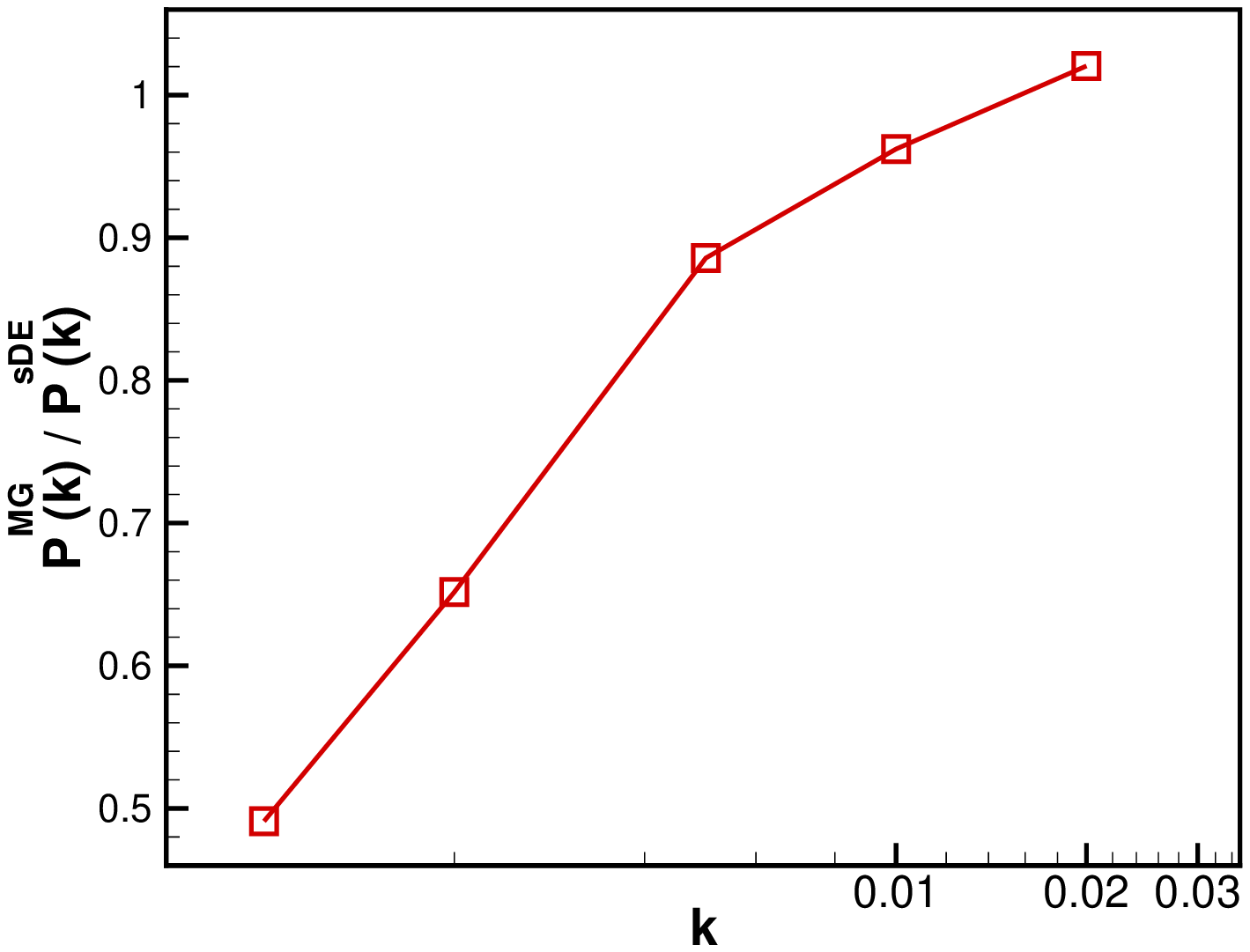,width=300pt}  \caption{The
fraction of MG to sDE power spectrum is plotted versus wavenumbers
at the present epoch.} \label{fig9} }

In order to compare power spectrum  predicted from the equivalence
DE and MG models, we obtain the density contrast in DE model for
various wavenumbers at the present time. We note that the only
difference in the differential equation governing the density
contrast evolution of the MG and DE is that, for the case of DE we
let $Q,\gamma = 1$. We plot the fraction of the power spectrum of
the matter density in the MG to that in the equivalence DE in terms
of wavenumber at present time in Fig.(\ref{fig9}). While we have
equivalent dynamics for the background expansion of the universe,
the power spectrum from the MG and DE models are different. For the
structures with the size of $k < 0.01$ the MG power spectrum is
smaller than the DE model, as the screened mass in these scales is
less than unity. On the other hand for structures with the scale of
$k> 0.01$ the screened mass is larger than unity and the ratio of
the two power spectrums increases. The dependence of the power
spectrum to the wavenumber can provide an observational tool to
distinguish between these two models. Qualitatively we can also
assert that as the screened mass function in small scales  is bigger
than unity, we will have an enhancement in mater density power
spectrum  in comparison to the equivalent sDE model.

Another observational parameter in the structure formation is the
dynamics of the growth index factor defined as $f={d\ln\delta}/{d\ln
a}$ in terms of the redshift. We rewrite the differential Eq.(
\ref{Eqdcredshift}) in terms of growth index factor $f$ as below:
\begin{equation} \label{Eqgrowth}
f^{'}-\frac{f^2}{1+z}+(A-\frac{1}{1+z})f-B(1+z)=0,
\end{equation}
\FIGURE{\epsfig{file=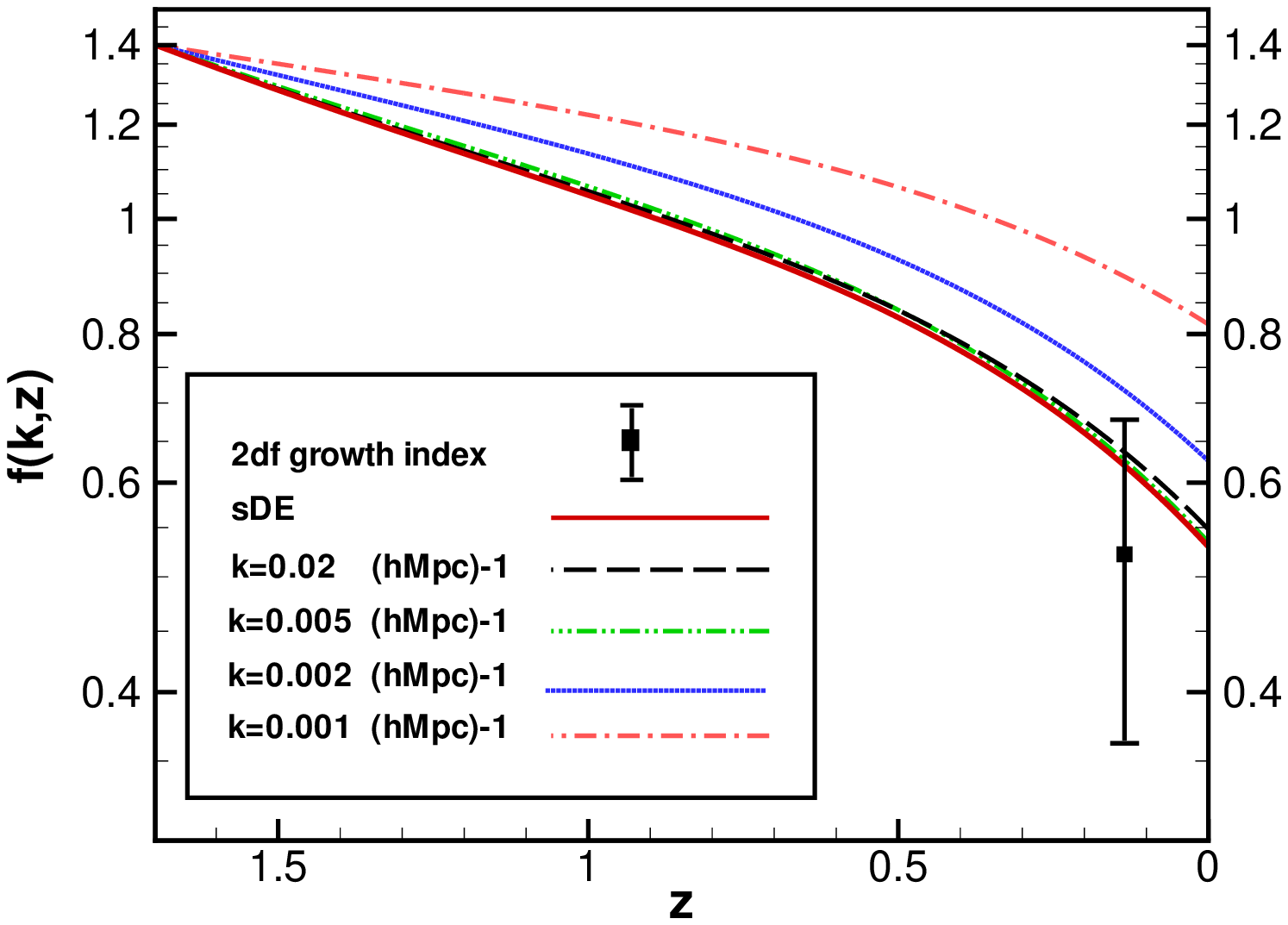,width=300pt}  \caption{The growth
index function is plotted versus redshift for various wavenumbers.
The observational data from the 2dF survey is represented with the
corresponding error bar. } \label{fig10}}
where A and B are defined
from Eqs. (\ref{A}) and (\ref{B}). We plot the growth index in the
MG and sDE in Fig.(\ref{fig10}) with the observational data from the
2dF survey \cite{2df}.  For the higher redshifts the growth index
for the two models is the same while for the later times they
diverge. An observational way to distinguish the MG from the DE
model is measuring the growth index for various wavenumbers. For the
case of DE model, the growth index is independent of the size of
structure, as the structure formation equation for the scales larger
than the Jeans length is independent of the $k$ while in the MG
model, screened mass relates the growth index of the structure to
its size. Future surveys of the large scale structure may reveal the
growth index in terms of wavenumber of the structures.

\section{ISW as probe to distinguish between DE and MG models}\label{section5}

In this section we examine the effect of MG on Integrated
Sachs-Wolfe effect (ISW) \cite{Sachs67} and compare it with the
dynamically equivalent dark energy model. ISW effect is one of the
important cosmological probes which traces the effect of dynamics of
the potentials on the photons of the cosmic microwave background
radiation (CMB). The effect of MG on ISW effect has already been
studied in several works in metric formalism as in \cite{Song2007}.
This effect has been also studied in a class of $f(R)$-gravities in
Palatini formalism where the deviation from $\Lambda$CDM of this
models occurs at a higher redshift (known as the early time f(R)
gravities)\cite{Zhang2006} and the results are compared with
ordinary $f(R)$ gravities. It was shown that the matter power
spectrum is more sensitive to gravity modification than the
ISW-effect, which is compatible to our results obtained in this
section, although their specific choice of gravity action caused to
a deviation in ISW effect in higher l-moments \cite{Baouji2006}.

We start with the reconstruction of the perturbed potential $\Phi$
which is a relevant parameter in the ISW and weak lensing
observations. Using the Poisson equation of (\ref{EqScreenedMass})
for MG, $\Phi$ is given in terms of the density contrast and
screened mass function as follows:
\begin{equation} \label{EqPhi}
\Phi(k,z)=\frac{3\Omega_{m}^{0}(1+z)}{2{\tilde{k}}^2}\delta_{m}Q(k,z).
\end{equation}

\FIGURE{ \epsfig{file=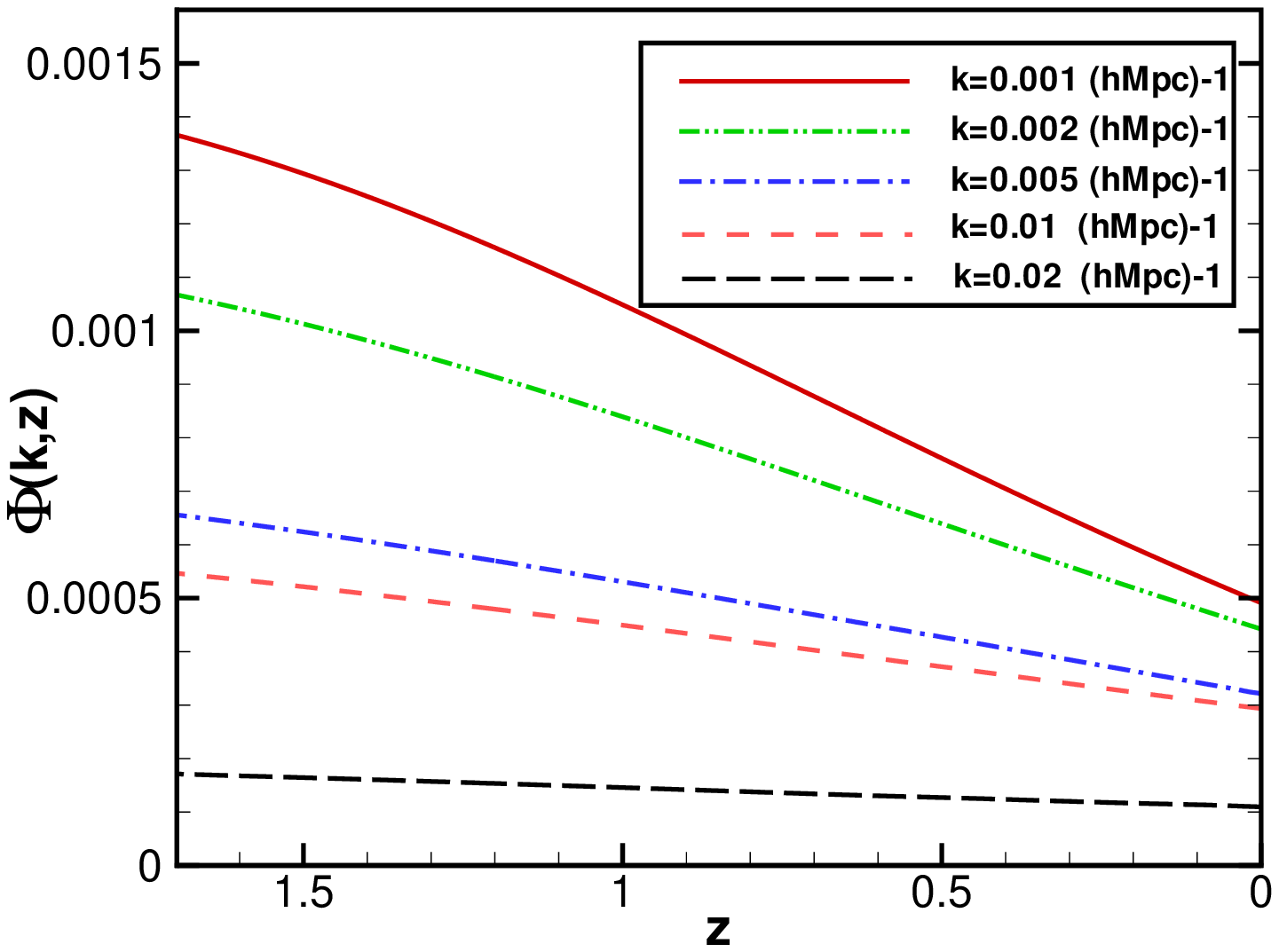,width=300pt}  \caption{The
perturbed potential $\Phi$ in MG theory in terms of redshift for
various wavenumbers.}\label{fig11}}

\FIGURE{ \epsfig{file=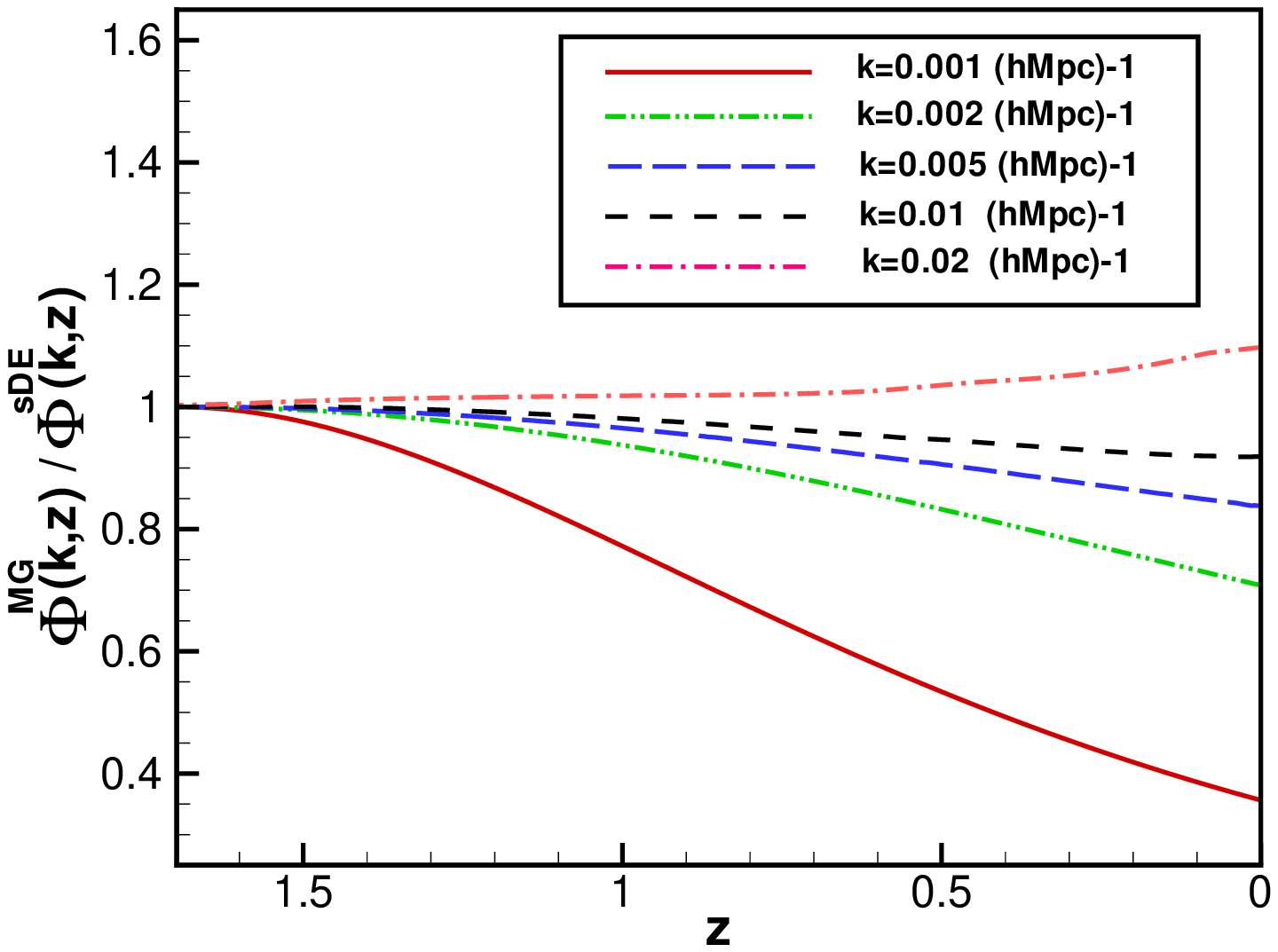,width=300pt}  \caption{The
fraction of perturbed potential in MG to that in the sDE model in
terms of redshift for various wavenumbers.}\label{fig12} }

Substituting the numerical values of matter density contrast and the
screened mass function, we obtain the potential in terms of redshift
for various wavenumbers. The result is plotted in Fig.(\ref{fig11}).
For comparison with the dark energy model, we express the relative
magnitude of Potential in MG and sDE models as below:
\begin{equation}\label{EqrelPhi}
\frac{\Phi^{MG}(k,z)}{\Phi^{sDE}(k,z)}=\frac{\delta^{MG}_{m}}{\delta^{sDE}_m}Q(k,z).
\end{equation}
Fig.(\ref{fig12}) shows this ratio in terms of redshift for various
wavenumbers. Eq.({\ref{EqrelPhi}}) shows that this ratio is
proportional to the relative magnitude of density contrast and
screened mass function. In the zeroth order the density contrast
growth is approximately equal to density contrast in sDE. The
relative magnitude of potentials diverge from unity in large scales
as the screened mass function k-dependance. Although the
k-dependence of $\gamma$-parameter has a secondary effect on the
growth of density contrast through Eq.(\ref{Eqdcredshift}).

Now we apply results from the dynamical evolution of potential in
the ISW effect. ISW effect is caused by the time variation in the
cosmic gravitational potential, as the CMB photons pass through the
structures. In this effect the amount of the energy gain of the
photons due to the gravitational potential of the structure may not
be equal to the energy lose of the photons when they leave the
potential well. The result of this effect is an extra anisotropy on
the CMB map.

In generic case, the CMB anisotropy due to the ISW is given by
\begin{equation} \label{EqModTheta}
\Theta_{l}(k,\eta_{0})=\int_{0}^{\eta} d\eta
e^{-\tau}\left[\dot{\Psi}(k,\eta)-\dot{\Phi}(k,\eta)\right]j_{l}(k(\eta_{0}-\eta)),
\end{equation}
where $\Theta_{l}$ represents the l-th moment of temperature
contrast defined in Eq.(\ref{EqTheta}), $\tau$ is the optical depth
of the photons which we can neglect it in ISW calculations, $j_l$ is
the Bessel function and $\eta$ is the conformal time and "$.$" in
this equation represents derivative with respect to the conformal
time. Since in the smooth dark energy models $\Phi=-\Psi$ Eq.
(\ref{EqModTheta}) reduces to:
\begin{equation}\label{EqQUTheta}
\Theta_{l}^{sDE}(k,\eta_{0})=\int_{0}^{\eta}
d\eta\left[-2\dot{\Phi}^{DE}(k,\eta)\right]j_{l}(k(\eta_{0}-\eta)),
\end{equation}
superscript "$sDE$" on $\Theta$ indicates ISW effect in sDE model.
For calculating the ISW in the equivalent dark energy model, we use
the numerical value for the $\Phi$ which is obtained in the last
section. On the other hand, since in the MG, the two potentials of
$\Phi$ and $\Psi$ are not equal and relates by the slip parameter in
Eq. (\ref{slipf}), we eliminate $\Psi$ in favor of $\Phi$ in
Eq.(\ref{EqModTheta}) and rewrite the ISW effect in the MG models as
follows:
\begin{equation}
\Theta_{l}^{MG}(k,z)=\int_{z\sim 1100
}^{z=0}dz\frac{d}{dz}\left[-\Phi^{MG}(\gamma^{-1}+1)\right]j_{l}[\tilde{k}\int_{0}^{z}\frac{dz^{'}}{E(z)^{'}}],
\end{equation}
where the superscript of "$MG$" in $\Theta_{l}$ indicates the
temperature anisotropy in the MG model. For simplicity in the
numerical calculation, we reexpress time derivatives in terms of
redshift and since $\Phi$ is almost a constant function beyond the
redshift of $z \simeq 1.7$ (see Fig. \ref{fig11}), we put a cutoff
in the numerical integration up to this redshift.

In order to express our results in terms of observable parameter, we
obtain the $\Theta_{l}$ in terms of CMB power spectrum, $C_l$. The
variance of temperature fluctuations,
$\delta_{ll^{'}}\delta_{mm^{'}}C_{l}=<a_{l^{}m^{}}a^{*}_{l^{'}m^{'}}>$
can be related to $\Theta_{l}$ as below \cite{Dodelson}:
\begin{equation}\label{EqCl}
C^{ISW}_{l}=\frac{2}{\pi}\int_{0}^{\infty}dkk^2P(k)|\frac{\Theta_{l}(k)}{\delta(k)}|^2.
\end{equation}

\FIGURE{ \epsfig{file=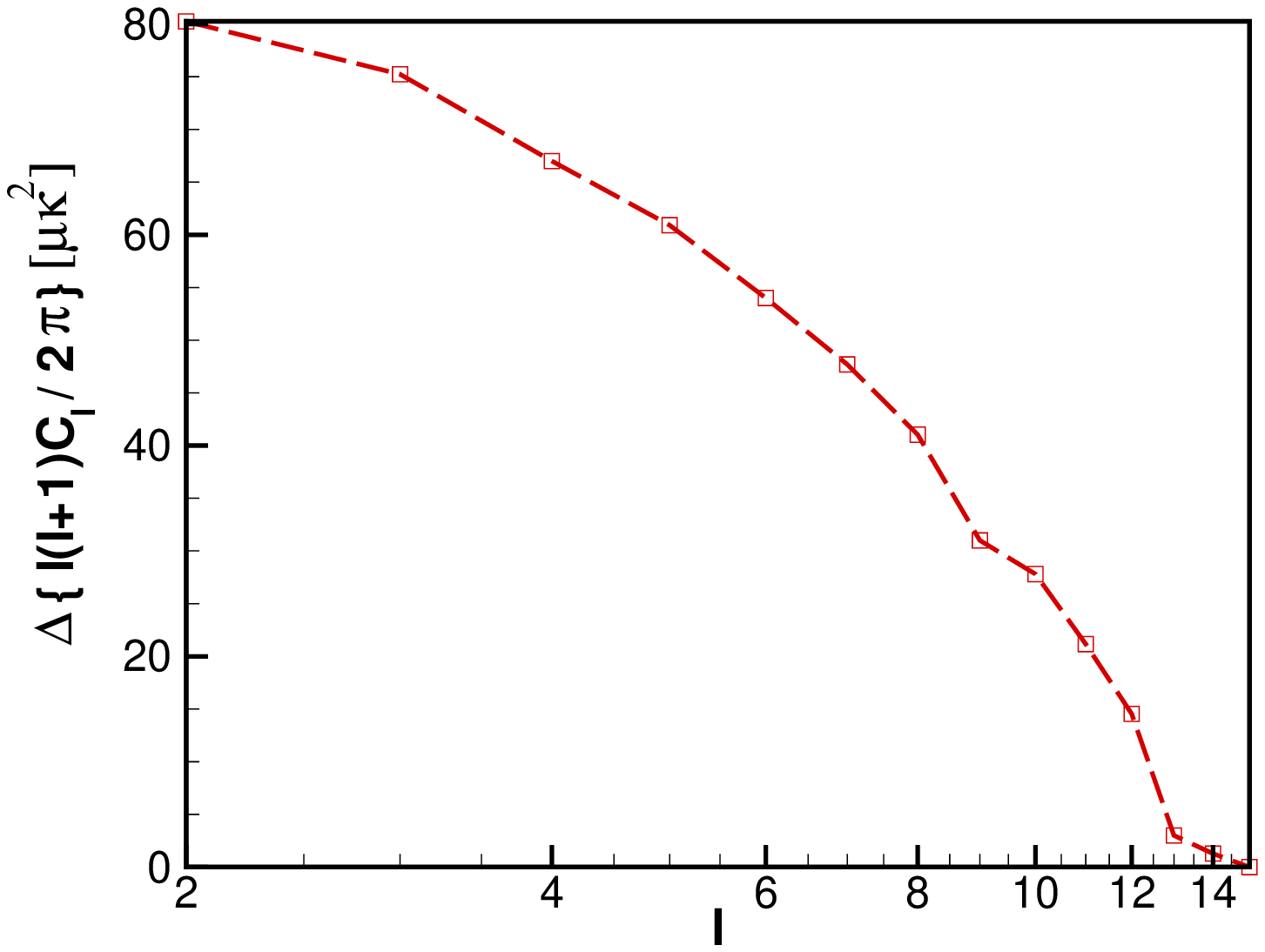,width=300pt}  \caption{The
difference of $\Delta C_{l}=C^{sDE}_{l}-C^{MG}_{l}$ between the
equivalent DE and MG models in terms of $l$.} \label{fig13} }

We note that for the small scale structures, the time derivative of
the potential in Fig. (\ref{fig11}) is almost zero. Hence we expect
that in the ISW, small structures ( large $l$s ) will not be
modified. So we examine only small moments for calculating deviation
from the standard CMB power spectrum. Taking small $l$s in the range
of $l<10$ we calculate $C_{l}$s both in the MG and the equivalent DE
model. The numerical difference of the power spectrum $\Delta
C_{l}=C^{sDE}_{l}-C^{MG}_{l}$ between the DE and MG models in terms
of $l$ is shown in Fig.(\ref{fig13}). Here for the small $l$s we
have more contrast than the large $l$s.

\FIGURE{ \epsfig{file=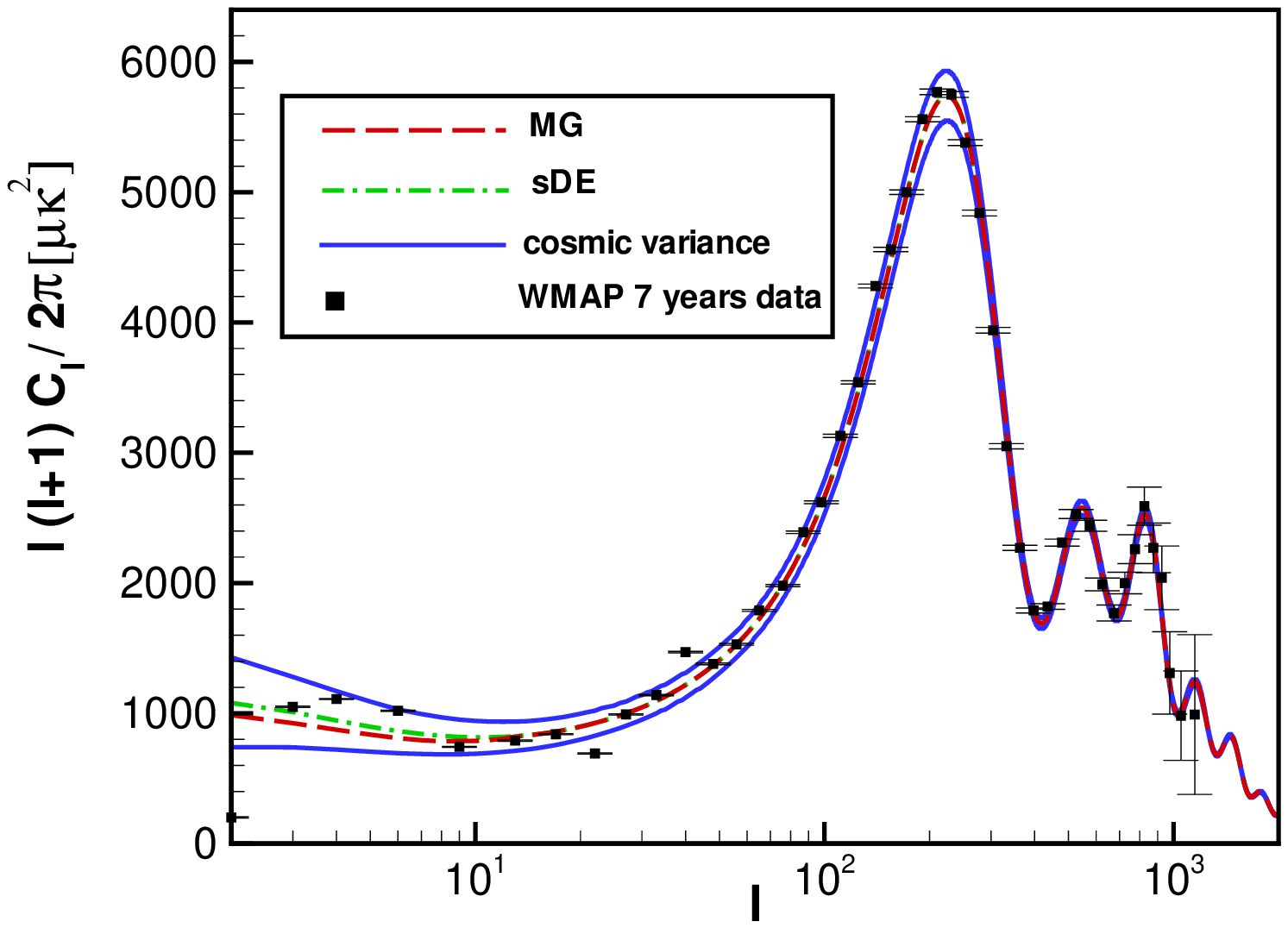,width=300pt}  \caption{Power
Spectrum of CMB for Modified gravity and equivalent dark energy
model in the Palatini formalism.} \label{fig14}}

Finally we express our analysis in CMB power spectrum for the MG and
sDE models and compare them with the observational data. We use the
CMB-fast code to plot the power spectrum for the equation of state
defined in Eq.(\ref{EqCPL}). As we showed in Fig. (\ref{fig13}), the
difference in the power spectrum between the two equivalent DE and
MG models is in the small $l$s and for the larger $l$s the power
spectrum of the two models coincide to each other. For smaller $l$s,
we use the numerical results of Fig. (\ref{fig13}) to obtain the
power spectrum of the MG from the DE model. Fig.(\ref{fig14})
compares the power spectrum from the MG and DE model with the latest
WMAP-7 years data \cite{Jar2010}. We note that this deviation is
smaller than the cosmic variance in the WMAP data.

\section{Conclusion}

\label{conc}

One of the important questions in the problem of the acceleration of
the universe is either this acceleration is produced by a dark
energy fluid or that is a manifestation of the gravity modification.
It has been shown that the expansion history of the universe can not
dynamically distinguish between these two models and at this stage
they are equivalent.

In this work we compared various  aspects of the structure formation
in the Modified Gravity (MG) and smooth Dark Energy (sDE) models. We
used an ansatz for the equation of state of a dark energy model and
obtained the equivalent modified gravity . Using a synthetic SNIa
data set from the SNAP observation, we did the inverse problem
approach and obtained the effective action of the gravity (i.e.
$f(R)$). It has been shown that the structure formation unlike to
the background dynamics can distinguish between these two models.

In the modified gravity, the modification of the Einstein-Boltzman
equation for the structure formation has extra terms compare to the
sDE model. This modification can be given in terms of screened mass
function and gravitational slip parameter. We obtained these two
parameters in the MG model and solved the differential equation for
the evolution of the structure formation. Using the
Harrison-Zeldovich initial condition for the structures, finally we
obtained the power spectrum and the growth index of the structures
in the two scenarios in the linear regime of the structure
formation. We have shown that for the structures with the larger
size, the effective spectral index at the present time is slightly
more than one. From the observational point of view, sampling of
large scale structure more than $100$Mpc is needed to test this
effect. On the other hand, we showed that the growth index parameter
in the MG unlike to the sDE models is a scale dependent parameter.
Next generation of peculiar velocity survey may measure this effect.
Finally we obtained the effect of the MG structure formation on the
Integrated Sachs-Wolfe effect on CMB map. We showed that the
deviation of the CMB power spectrum due to the MG is larger in the
small $l$s, however it is not too large to distinguish between the
two models due to the uncertainty from the cosmic variance.

\appendix
\section{Structure formation probes in metric formalism}

Here in this appendix, we obtain the evolution of matter density
contrast in the metric formalism and consequently derive the matter
power spectrum and growth index in order to compare them with that
in the Palatini formalism.

\FIGURE{ \epsfig{file=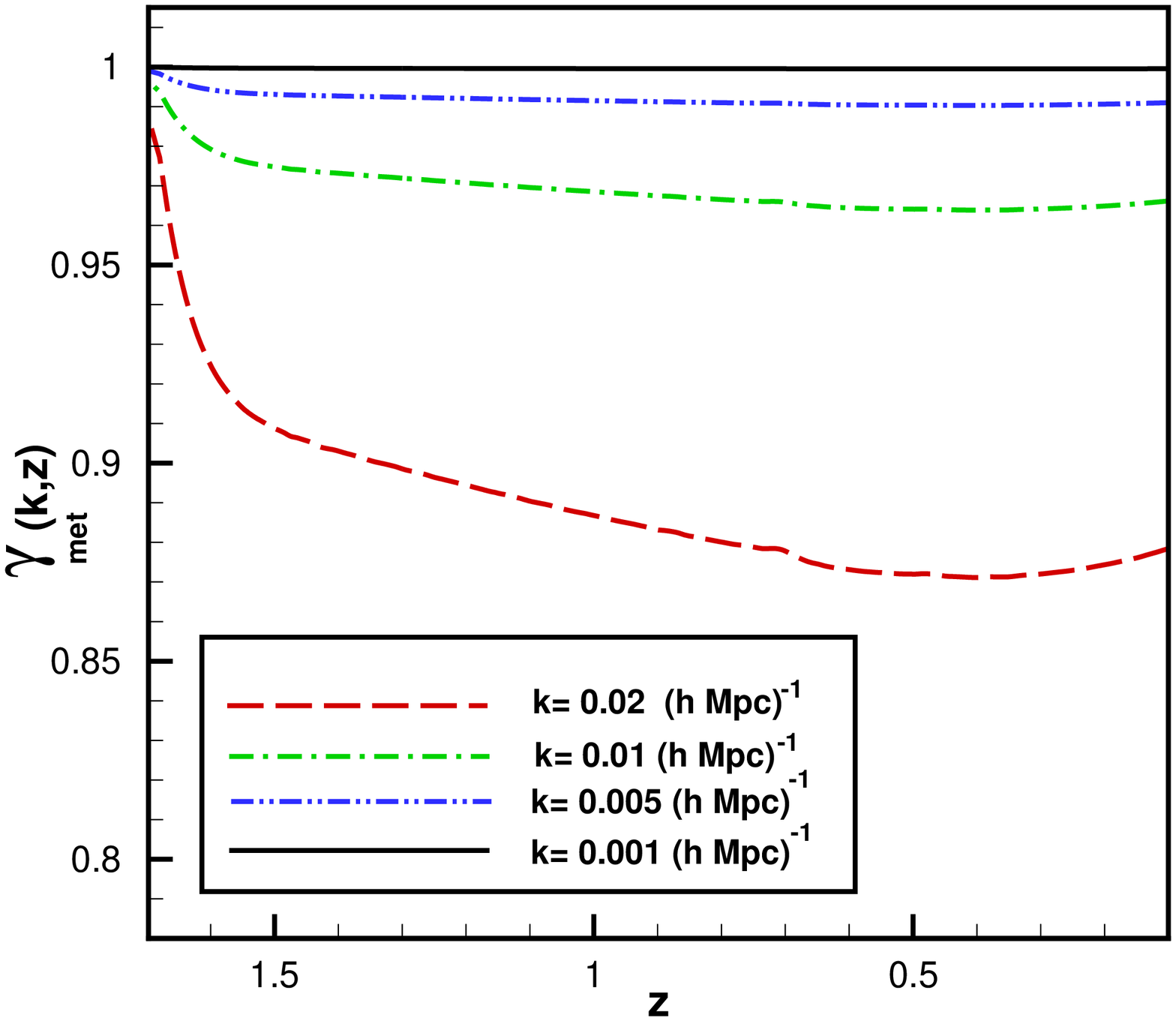,width=300pt}[h]  \caption{The
gravitational slip parameter in metric formalism for different
wavenumbers. The Long dashed line correspond to small scales
$k=0.02$, where we have the most deviation in this parameter from GR
value.} \label{Fig-g-met}}

\FIGURE{ \epsfig{file=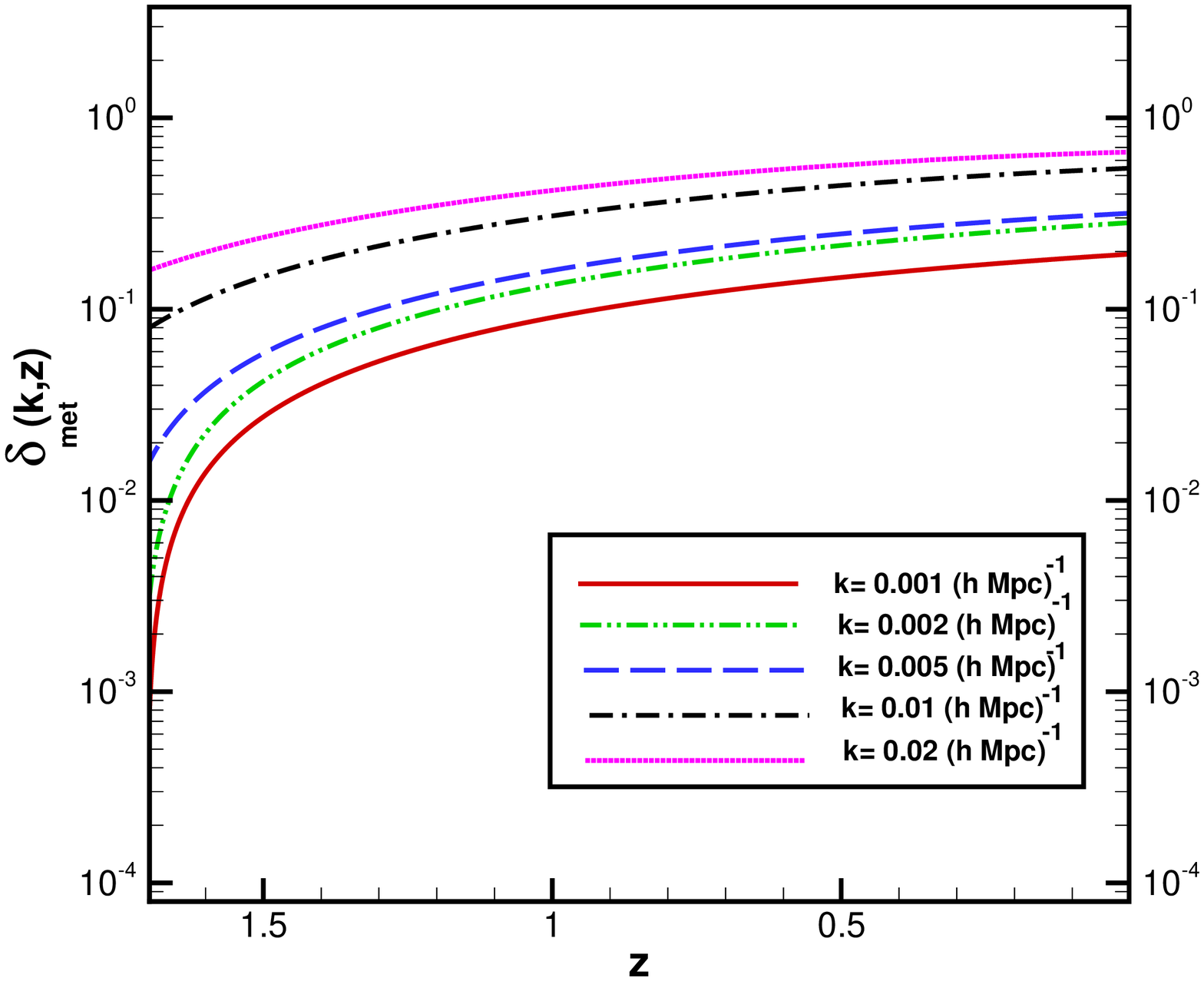,width=300pt} \caption{The
density contrast of the mater in metric formalism in terms of
redshift and wavenumber is plotted. parameter in metric formalism
for different wavenumbers.} \label{fig-delta-metric}}

\FIGURE{ \epsfig{file=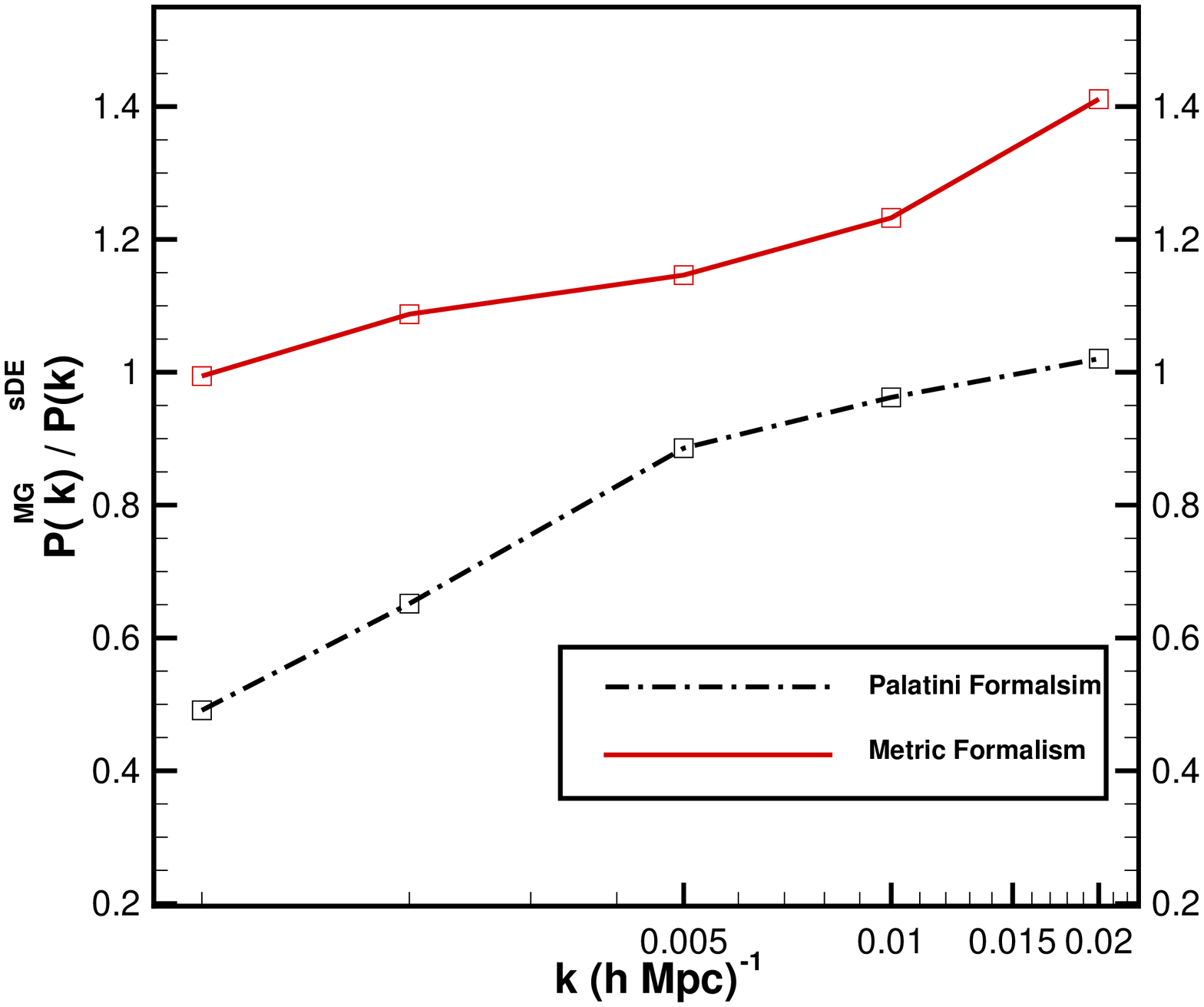,width=300pt} \caption{The
relative power Spectrum of Modified gravity and smooth Dark Energy
is plotted versus wavenumber for Metric formalism(solid line),
Palatini Formalism (dash-dot line)} \label{power-rel-met-pal}}


 \FIGURE{\epsfig{file=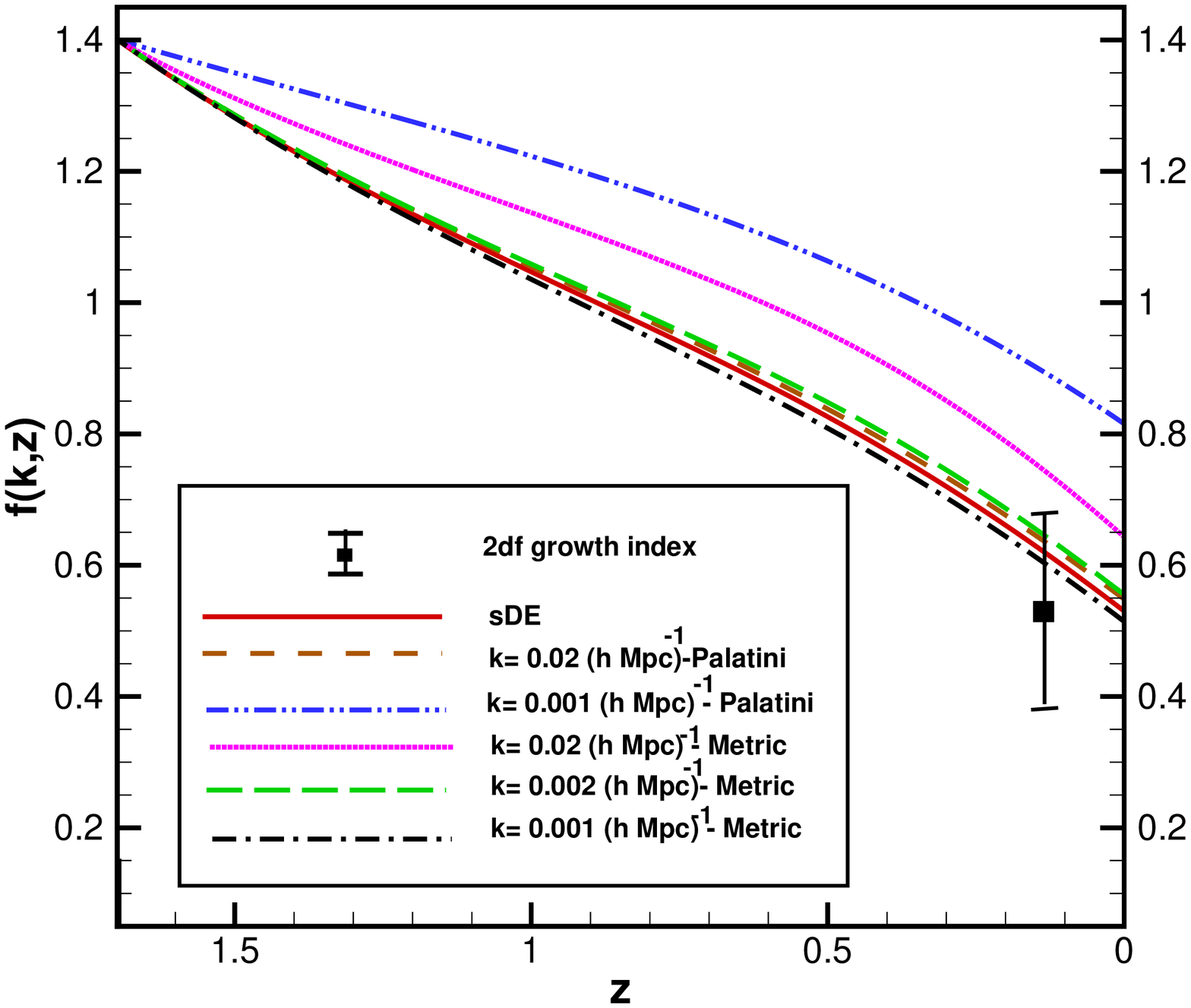,width=300pt} \caption{The growth index
factor is plotted versus redshift for different wavenumbers in
Metric formalism (dotted, long dash, dash-dot lines), Palatini
Formalism (dashed and dash-dot-dot lines) and sDE models(solid
line)} \label{f-rel}}

 \FIGURE{\epsfig{file=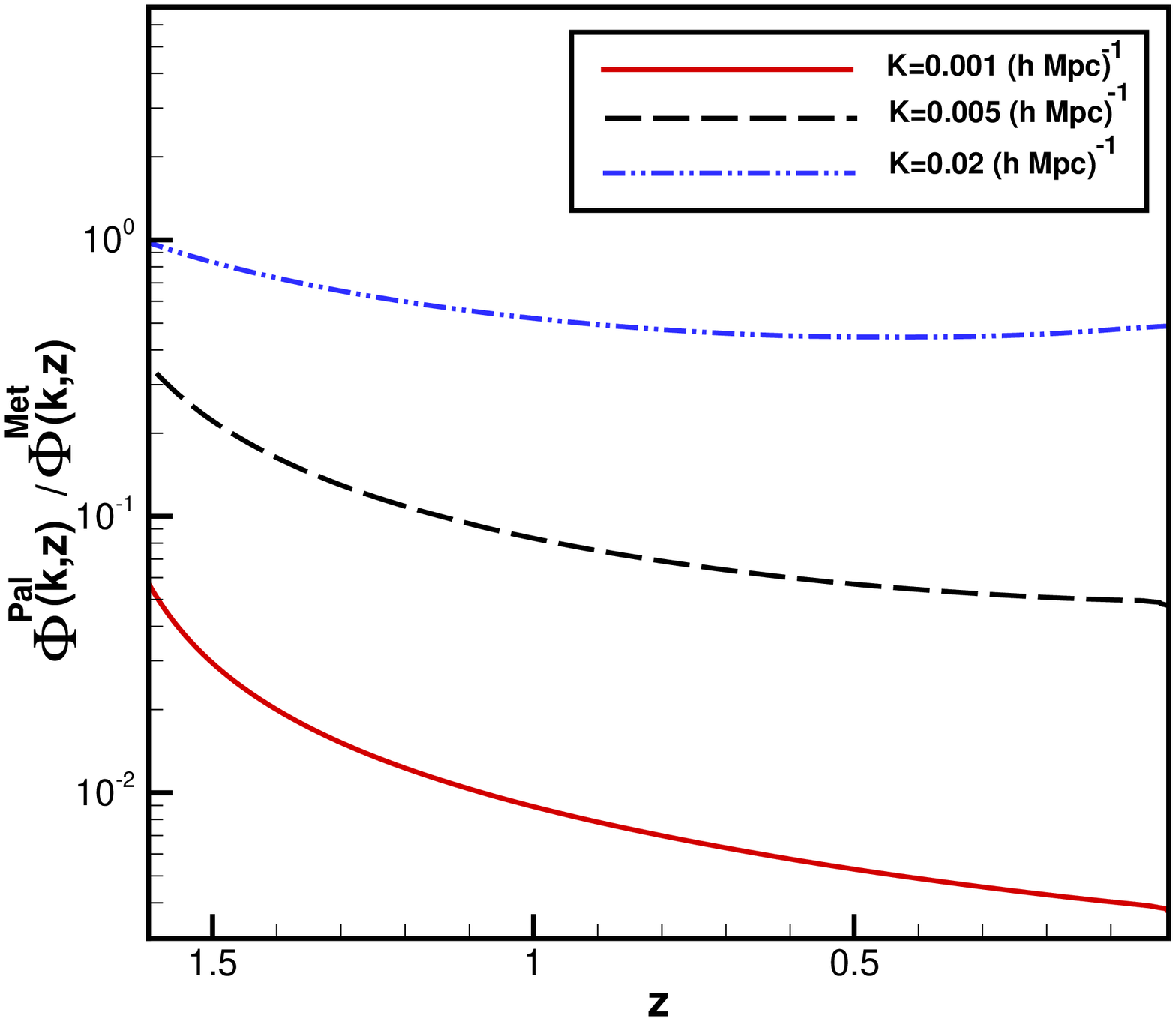,width=300pt} \caption{The relative
 magnitude of gravitational potential in Metric and Palatini formalism plotted versus
 redshift for different wavenumbers} \label{Phi-met-pal}}

In order to determine the matter density evolution we use
Eq.(\ref{EqMetricDensity}), where we can  rewrite it, in terms of
redshift as below:
\begin{equation}\label{EqMetricDensityRedshift}
\delta^{''}+\left[\frac{E^{'}}{E}-\frac{1}{1+z}\right]\delta^{'}-\frac{3}{2}\frac{\Omega^{0}_{m}(1+z)}{E^2}\gamma_{met}^{-1}Q_{met}\delta=0.
\end{equation}
The above equation has the same form as in the Palatini formalism,
Eq.(\ref{EqPAlatiniDensityRedshift}), however the difference is in
the value of the screened mass function and the gravitation slip
parameter in the two modified gravity formalism.

The screened mass function is obtained in Sec.(\ref{section4}) and
plotted in Fig.(\ref{fig-metric}). For calculating the gravitational
slip parameter, due to the slow change of the action over the Hubble
time we neglect the term of $\delta\Box F(R)$ in the perturbation of
the trace of the field equation (\ref{metricTrace}). We follow the
same procedure an in the Palatini formalism and obtain the
gravitational slip parameter in terms of reconstructed $F$ from
Eq.(\ref{Fmetric}) and the screened mass $Q_{met}$ from
Eq.({\ref{Q-metric}}) which results in
\begin{equation}
\gamma_{met}=\left[1-\frac{2m(1+z)^2{\tilde{k}}^2}{\tilde{R}FQ_{met}(m-1)}\right]^{-1}.
\end{equation}
The gravitational slip parameter in the metric formalism is Plotted
in Fig.(\ref{Fig-g-met}) for different wavenumbers. Similar to the
Palatini case, Fig.(\ref{fig6}), the gravitational slip parameter
deviates from its GR value $\gamma=1$ in small scales. Knowing the
screened mass function and gravitational slip parameter, we solve
the differential Eq.({\ref{EqMetricDensityRedshift}}) governing the
evolution of density contrast. The results is plotted in
Fig.(\ref{fig-delta-metric}).

In the next step we compare the relative magnitude of power spectrum
of metric-MG to sDE, $P_{Met}/P_{sDE}$ and Palatini-MG to sDE ,
$P_{Pal}/P_{sDE}$  both plotted in Fig.({\ref{power-rel-met-pal}}).
As we except because the variation of screened mass function from GR
value $Q>1$ in small scales, we get a amplification in power
spectrum in large $k$'s, which shows a different prediction than the
Palatini formalism in which there is a suppression in large scales.

Using Eq.({\ref{EqMetricDensityRedshift}}) we can also reexpress the
growth index function $f\equiv\frac{d\ln \delta}{d\ln a}$ in terms
of redshift as:
\begin{equation}
f^{'}-\frac{f^2}{1+z}+\left[\frac{E^{'}}{E}-\frac{2}{1+z}\right]f+\frac{3}{2}\frac{\Omega_{m}^{0}}{E^2}\gamma^{-1}_{met}Q_{met}(1+z)^2=0
\end{equation}
 For comparison with the Palatini
formalism, we plot the growth index for large and small scales in
both metric and Palatini formalism in Fig.(\ref{Phi-met-pal}).\\

Finally the last comparison is done for the relative magnitude of
gravitational potential $\Phi$ in both formalism. From Poison
equation, it is straightforward to calculate this ratio as:
\begin{equation}
\frac{\Phi_{Pal}}{\Phi_{Met}}(k,z)=\frac{\delta^{Pal}_{m}Q_{Pal}}{\delta^{Met}_{m}Q_{Met}}
\end{equation}
The results for three wavenumbers is plotted in
Fig.(\ref{Phi-met-pal}). The relative magnitude of gravitational
potential in higher-redshifts converge to unity for all wavenumbers.
This is because the screened mass functions, in both formalism
converge to unity and the evolution of matter density contrast in
deep matter dominated era is proportional to the scale factor.

\end{document}